%% file: quant-ph0007060
\documentclass[12pt]{article}
\usepackage{amssymb}
\usepackage{amsfonts}
\usepackage{amsthm}
\usepackage{latexsym}
\input diagram.tex %

\newtheorem{thm}{Theorem}
\newtheorem*{rsthm}{Reeh-Schlieder Theorem}
\newtheorem*{fullrs}{Full Reeh-Schlieder Theorem}

\newtheorem*{lemma}{Lemma}
\theoremstyle{definition}
\newtheorem*{defn}{Definition}

\newcommand{\norm}[1]{\mbox{$\| #1\|$}}
\newcommand{\alg}[1]{\mathfrak{#1}}
\newcommand{\hil}[1]{\mathcal{#1}}

\title{Reeh-Schlieder Defeats Newton-Wigner: \\ On alternative
  localization schemes in relativistic quantum field theory }
\author{Hans Halvorson \\
{\small \emph{Department of Philosophy, University of Pittsburgh} } \\
{\small \emph{E-mail: hphst1+@pitt.edu}} }
\date{\today}
\begin{document}
\maketitle 
\begin{abstract}  Many of the ``counterintuitive'' features of relativistic 
  quantum field theory have their formal root in the Reeh-Schlieder
  theorem, which in particular entails that local operations applied
  to the vacuum state can produce any state of the entire field.  It
  is of great interest then that I.E. Segal and, more recently, G.
  Fleming (in a paper entitled ``Reeh-Schlieder meets Newton-Wigner'') have proposed an
alternative ``Newton-Wigner'' localization scheme that avoids the Reeh-Schlieder theorem.  In this
paper, I reconstruct the Newton-Wigner localization scheme and clarify the \emph{limited} extent
to which it avoids the counterintuitive consequences of the Reeh-Schlieder theorem.  I also argue
that neither Segal nor Fleming have provided a coherent account of the physical meaning of
Newton-Wigner localization.  \end{abstract}

\section{Introduction} 
Relativistic quantum theory presents us with us a set of peculiar
interpretive difficulties over and above the traditional ones of
elementary quantum mechanics.  For example, while the notion of a
``localized object'' has a transparent mathematical counterpart in
elementary quantum mechanics, it appears that not every aspect of our
common-sense notion of localization can be maintained in the context
of relativistic quantum theory (cf. Malament 1996).  Many of the
thorny issues involving localization in relativistic quantum
\emph{field} theory have a common formal root in the so-called
``Reeh-Schlieder theorem.''  Thus, it is of particular philosophical
interest that I.E. Segal (1964) and, more recently, G. Fleming (2000)
have claimed to be able to avoid the Reeh-Schlieder theorem---and
thereby its ``counterintuitive'' consequences---by means of a
judicious reworking of the standard association between observables
and regions of space.\footnote{Saunders 1992 provides an extensive
  discussion of Segal's approach, although with different points of
  emphasis than the current presentation.  Although Fleming (1996, 12)
  appears to dismiss Saunders' comparison of his and Segal's
  approaches, Fleming's proposal for NW-local algebras (Fleming 2000)
  (prior to ``covariant generalization'') is mathematically identical
  to Segal's (1964) proposal.}

I am not convinced, however, that Segal and Fleming's
``Newton-Wigner'' localization scheme offers any satisfying resolution
for the ``problem'' of localization in relativistic quantum field
theory.  In particular, the Newton-Wigner localization scheme is not
completely immune from the consequences of the Reeh-Schlieder theorem;
and neither Segal nor Fleming has offered a conceptually coherent
description of the physical meaning of Newton-Wigner localization.

The context of the Reeh-Schlieder theorem is the axiomatic (or
algebraic) approach to quantum field theory.  This approach singles
out a family of postulates that apply quite generally to ``physically
reasonable'' quantum field models, and these postulates are used as a
starting point for further structural investigations.  One might
expect, then, that Segal and Fleming would attempt to undercut the
Reeh-Schlieder theorem by questioning one of the assumptions it makes
concerning which models are ``physically reasonable.''  However, Segal
and Fleming do not discuss the Reeh-Schlieder theorem at this level of
generality; rather, their discussion of the Reeh-Schlieder theorem is
restricted to a concrete field model, viz., the free Bose field.

I begin then in Section~2 with a brief review of the global structure
of the free Bose field model.  In Section~3, I present the standard
recipe for assigning observables to regions in space, and I explicate
the counterintuitive consequences---stemming from the Reeh-Schlieder
theorem---of this standard localization scheme.  In Section~4, I
present the Newton-Wigner localization scheme and show how it
``avoids'' the counterintuitive consequences of the Reeh-Schlieder
theorem.  Finally, in Sections~5 and~6, I argue that Reeh-Schlieder
has the final word against the Newton-Wigner localization scheme.

\section{The free Bose field} 
In this section, I briefly review the mathematical formalism for the
quantum theory of the free Bose field.  Although my presentation
differs from Fleming's (2000) in being more abstract and in its
emphasis on mathematical rigor, I take it that all parties agree
concerning the \emph{global} structures of the free field model (at
least in the absence of measurement interactions).  That is, we agree
on our answers to the following four questions:
\begin{enumerate}
\item What is the state space?
\item What are the observables (i.e., physical quantities)?
\item When no measurements are being made, how does the system evolve
  in time?  In other words, what is the (free) Hamiltonian?
\item What is the ground (i.e., vacuum) state?
\end{enumerate}
Disputes arise only at the level of the \emph{local} structure of the
free field model; e.g., which states are ``localized'' in this region
of space?  In this section, I spell out the answers to questions 1--4.
In Section~3, I take up questions concerning localization.

Recall that in its heuristic formulation, the free scalar quantum
field is described by an ``operator-valued field'' $\Phi$ on Minkowski
spacetime that solves the Klein-Gordon equation
\begin{equation} \frac{\partial ^{2}\Phi}{\partial t^{2}}+m^{2}\Phi =\nabla
  ^{2}\Phi ,\label{kg} \end{equation} and that satisfies the appropriate (equal-time)
canonical commutation relations.  As is well-known, however, there are mathematical
difficulties with understanding $\Phi$ as an operator-valued function.
A more rigorous approachtakes $\Phi$ as an ``operator-valued
distribution.''  That is, for each smooth, real-valued test-function $f$ on
Minkowski spacetime, $\Phi (f)$ can be defined as an operator on some Hilbert space.  

For my purposes here, it will be more convenient to turn to another
(mathematically equivalent) representation of the field $\Phi$.  Let
$C^{\infty}_{0}(\mathbb{R}^{3})$ denote the vector space of smooth,
compactly supported functions from $\mathbb{R}^{3}$ into $\mathbb{R}$,
and let
\begin{equation} S=C^{\infty}_{0}(\mathbb{R}^{3}) \oplus
  C^{\infty}_{0}(\mathbb{R}^{3}) .\end{equation}
Recall now that a scalar-valued solution $\phi$ of the Klein-Gordon
equation is uniquely determined by its Cauchy data (i.e., its
values, and the values of its first derivative) at any fixed time.
Thus, there is a one-to-one correspondence between elements of $S$ and
(a certain subset of) the space of solutions of the Klein-Gordon
equation.  Moreover, the
conserved four-vector current $\phi \stackrel{\leftrightarrow}{\partial} _{\mu}
\psi $ gives rise to a symplectic form $\sigma$ on $S$:
\begin{equation}
\sigma ( u_{0}\oplus u_{1},v_{0}\oplus v_{1})= \int
_{\mathbb{R}^{3}}(u_{0} v_{1}-u_{1}v_{0} )\, d^{3}\mathbf{x}
. \end{equation}  
We let $D_{t}$ denote the natural (inertial) symplectic flow on $S$;
i.e., $D_{t}$ maps the time-zero Cauchy data of $\phi$ to the 
time-$t$ Cauchy data of $\phi$.  The triple $(S,\sigma ,D_{t})$
contains the essential information specifying the classical theory of the scalar
field of mass $m$.  

A representation of the Weyl form of the canonical commutation
relations (CCRs) is a mapping $f\mapsto W(f)$ of $S$ into unitary
operators acting on some Hilbert space $\hil{K}$ such that $W(0)=I$
and \begin{equation} W(f)W(g) =e^{-i\sigma
    (f,g)}W(f+g).\end{equation} I will now sketch the construction
of the unique (up to unitary equivalence) ``Minkowski vacuum
representation'' of the CCRs.  This construction proceeds in two
steps.  In \emph{first quantization}, we ``Hilbertize'' the classical
phase space $S$, and we ``unitarize'' the classical dynamical group
$D_{t}$.  More precisely, suppose that $\hil{H}$ is a Hilbert space,
and that $U_{t}$ is a weakly continuous one-parameter group of
unitary operators acting on $\hil{H}$.  Suppose also that the
infinitesimal generator $A$ of $U_{t}$ is a positive operator; i.e.,
$(f,Af)\geq 0$ for all $f$ in the domain of $A$.  If there is a
one-to-one real-linear mapping $K$ of $S$ into $\hil{H}$ such that
\begin{enumerate}
\item $K(S)+iK(S)$ is dense in $\hil{H}$,
\item $2\mathrm{Im}(Kf,Kg)=\sigma (f,g)$,
\item $U_{t}K=KD_{t}$,
\end{enumerate}
then we say that the triple $(K,\hil{H},U_{t})$ is a
\emph{one-particle structure} over $(S,\sigma ,D_{t})$.  Constructing
a one-particle structure over $(S,\sigma ,D_{t})$ is a mathematically
rigorous version of ``choosing the subspace of positive frequency
solutions'' of the space of complex solutions to the Klein-Gordon
equation.

If there is a one-particle structure over $(S,\sigma ,D_{t})$, then it
is unique up to unitary equivalence (Kay 1979).  That is, suppose that
$(K,\hil{H},U_{t})$ and $(L,\widetilde{\hil{H}},\widetilde{U}_{t})$
are one-particle structures over $(S,\sigma ,D_{t})$.  Then, $L\circ K
^{-1}$ extends uniquely to a unitary mapping $V$ from $\hil{H}$ onto
$\widetilde{\hil{H}}$.
\begin{center} \resetparms \qtriangle[
  S`\mathcal{H}`\widetilde{\mathcal{H}};K`L`V]
\end{center}
It is also not difficult to see that $V$ intertwines the unitary
groups on the respective Hilbert spaces, i.e.,
$VU_{t}=\widetilde{U}_{t}V$.  This uniqueness result can be
interpreted as showing that the choice of time evolution in the
classical phase space suffices to determine uniquely the (first)
quantization of the classical system.

I will construct two (unitarily equivalent) versions of the
one-particle structure over $(S,\sigma ,D_{t})$.  First, we may
complete $S$ relative to the unique Hilbert space norm in which
time-evolution (given by $D_{t}$) is an isometry.  Specifically, let
$H$ denote the linear operator $(-\nabla ^{2}+m^{2})^{1/2}$ on
$C^{\infty}_{0}(\mathbb{R}^{3})$,\footnote{The mathematically rigorous
  definition of $H$ is as follows: Define the operator $A=-\nabla
  ^{2}+m^{2}$ on $C^{\infty}_{0}(\mathbb{R}^{3})$.  Then, $A$ is
  essentially self-adjoint, and the self-adjoint closure
  $\overline{A}$ of $A$ is a positive operator with spectrum in
  $[m^{2},\infty )$.  Using the functional calculus for unbounded
  operators, we may define $H=\overline{A}^{1/2}$, and it follows that
  the spectrum of $H$ is contained in $[m,\infty )$.}
and define a real inner-product $\mu$ on $S$ by
\begin{eqnarray} \mu (u_{0}\oplus u_{1},v_{0}\oplus
  v_{1})&=& (1/2)\left( ( u_{0},Hv_{0}) +(u_{1},H^{-1}v_{1}) \right) \\
  &=& (1/2)\left( \int _{\mathbb{R}^{3}}u_{0}(Hv_{0})\, d^{3}\mathbf{x}+
  \int
  _{\mathbb{R}^{3}}u_{1}(H^{-1}v_{1})\,d^{3}\mathbf{x} \right).\end{eqnarray}  
Now let $\hil{H}_{\mu}$ denote the completion of $S$ relative to the inner-product
$\mu$.\footnote{If $\mathcal{L}^{\pm }(\mathbb{R}^{3})$ denotes the
  completion of $C^{\infty}_{0}(\mathbb{R}^{3})$ relative to the inner
  product $(\,\cdot \,,H^{\pm 1}\,\cdot \,)$, 
then $\hil{H}_{\mu}=\mathcal{L}^{+}(\mathbb{R}^{3})\oplus
  \mathcal{L}^{-}(\mathbb{R}^{3})$.}   Define an operator $J$ on $\hil{H}_{\mu}$ by setting 
\begin{equation} J(u_{0}\oplus u_{1})= -H^{-1}u_{1}\oplus Hu_{0} \, ,
\end{equation}  
on the dense subset $S$ of $\hil{H}_{\mu}$.  Clearly $J^{2}=-I$, i.e.,
$J$ is a ``complex structure'' on $\hil{H}_{\mu}$.  Thus,
$\hil{H}_{\mu}$ becomes a complex vector space when we define scalar
multiplication by $(a+ib)f=af+J(bf)$, and is a complex Hilbert space
relative to the inner-product
\begin{eqnarray} (f,g) _{\mu}&=&\mu (f,g)+i\mu (Jf,g)  \\
&=& \mu (f,g)+(i/2)\sigma (f,g) .\end{eqnarray} 
Finally, it can be shown that $[J,D_{t}]=0$, so that $D_{t}$ extends uniquely to a weakly
continuous one-parameter group of \emph{unitary} operators (denoted again by
$D_{t}$) on the complex Hilbert space $\hil{H}_{\mu}$.  Therefore, $(\iota,
\hil{H}_{\mu},D_{t})$, with $\iota$ the identity mapping, is a 
one-particle structure over $(S,\sigma ,D_{t})$.  

It may not be immediately obvious---especially to those accustomed to non-relativistic quantum
mechanics---how to tie the physics of localization to the mathematical structure of the Hilbert
space $\hil{H}_{\mu}$.  (For example, which vectors in $\hil{H}_{\mu}$ are localized in a given
spatial region?)  The Newton-Wigner one-particle structure brings us back to familiar
territory by using the space $L^{2}(\mathbb{R}^{3})$ as the concrete
representation of the one-particle space.  In particular, define the
mapping $K:S\mapsto L^{2}(\mathbb{R}^{3})$ by
\begin{equation} K(u_{0}\oplus u_{1})\:=\:2^{-1/2}(
  H^{1/2}u_{0}+iH^{-1/2}u_{1} ) .\end{equation}  
It is then straightforward to check that the complex-linear span of $K(S)$ is dense in
$L^{2}(\mathbb{R}^{3})$, and that $K$ preserves (modulo a factor of
$2$) the symplectic form $\sigma$.  Moreover, it
can be shown that $K$ intertwines $D_{t}$ with the one parameter unitary group
$U_{t}=e^{-itH}$ on $L^{2}(\mathbb{R}^{3})$.  Therefore,
$(K,L^{2}(\mathbb{R}^{3}),U_{t})$ is a one-particle structure over $(S,\sigma ,D_{t})$.

Since $(\iota ,\hil{H}_{\mu},D_{t})$ and $(K,L^{2}(\mathbb{R}^{3}),U_{t})$ are one-particle
structures over $(S,\sigma ,D_{t})$, it follows that $(K\circ \iota \,^{-1})=K$
extends uniquely to a unitary operator $V$ from $\hil{H}_{\mu}$ onto
$L^{2}(\mathbb{R}^{3})$:
\begin{center}
  \resetparms \qtriangle[
  S`\mathcal{H}_{\mu}`L^{2}(\mathbb{R}^{3});\iota `K`V]
\end{center}
Thus, the one-particle spaces $(\hil{H}_{\mu},D_{t})$ and
$(L^{2}(\mathbb{R}^{3}),U_{t})$ are mathematically, and hence
physically, equivalent.  On the other hand, the two spaces certainly
\emph{suggest} different notions of localization.

\subsection{Second quantization}
Once we have a one-particle space $(\hil{H},U_{t})$ in hand, the
movement to a quantum \emph{field} theory (i.e., ``second
quantization'') is mathematically straightforward and uniquely
determined.\footnote{For a more detailed exposition, see Bratteli and Robinson 1997, Section
5.2.}  In particular, let $\hil{F}(\hil{H})$ denote the ``Fock
space'' over $\hil{H}$.  That is,
\begin{equation}
\hil{F}(\hil{H})=\mathbb{C}\oplus \hil{H} \oplus \hil{H}^{2} \oplus \hil{H}^{3} \oplus \cdots
,\end{equation}
where $\hil{H}^{n}$ is the $n$-fold symmetric tensor product of $\hil{H}$.   
As usual we let
\begin{equation} \Omega =1\oplus 0\oplus 0\oplus \cdots \end{equation} 
denote the vacuum vector in $\hil{F}(\hil{H})$.  For each $f\in
\hil{H}$, we define the creation $a^{+}(f)$ and annihilation $a(f)$
operators on $\hil{F}(\hil{H})$ as usual, and we let $\Phi (f)$ denote
the self-adjoint closure of the unbounded operator
\begin{equation}
2^{-1/2}(a(f)+a^{+}(f)) .\label{kraftwerk} \end{equation} 
If we let $W(f)=\exp \{ \,i\Phi (f)\}$, then the $W(f)$ satisfy the 
Weyl form of the canonical commutation
relations: \begin{equation} W(f)W(g)=e^{-i\,\mathrm{Im}(f,g)/2}W(f+g)
  ,\label{weyl} \end{equation} and vacuum expectation values are given
explicitly by
\begin{equation} \langle \Omega ,W(f)\Omega \rangle = \exp \left(
    -\norm{f}^{2} /4 \right) .\label{vacuum}
\end{equation}
The dynamical group on $\hil{F}(\hil{H})$ is given by the ``second
quantization'' $\Gamma (U_{t})=e^{itd\Gamma (H)}$ of the dynamical
group $U_{t}=e^{itH}$ on $\hil{H}$, and the vacuum vector $\Omega$ is the unique
eigenvector of the Hamiltonian $d\Gamma (H)$ with eigenvalue $0$.

\section{Local algebras and the Reeh-Schlieder theorem}
To this point we have only discussed the global structure of the free
Bose field model.  The physical observables for the free Bose field
are given by the self-adjoint operators on Fock space
$\hil{F}(\hil{H})$.  We equip this model with a \emph{local structure}
when we define a correspondence between regions in space and
``subalgebras'' of observables.  This labelling may be done for
various purposes, but the traditional motivation was to indicate those
observables that can (in theory) be measured in that region of space.

Now, each real-linear subspace $E$ of the one-particle space $\hil{H}$
gives rise naturally to a subalgebra of operators, viz., the algebra
generated by the Weyl operators $\{ W(f):f\in E\}$.  Thus, a
localization scheme needs only to determine which real-linear subspace
of $\hil{H}$ should be taken as corresponding to a region $G$ in
physical space.  \emph{It is on this point that the Newton-Wigner
  localization scheme disagrees with the standard localization
  scheme.}  In the remainder of this section, I discuss the standard localization scheme and its
consequences. 

The standard localization scheme assigns to the spatial region $G$ the
subset $S(G)\subseteq \hil{H}_{\mu}$ of Cauchy data localized in $G$.
That is, if $C^{\infty}(G)$ denotes the subspace of
$C^{\infty}_{0}(\mathbb{R}^{3})$ of functions with support in $G$,
then
\begin{equation} S(G)=C^{\infty}(G) \oplus C^{\infty}(G)
  ,\end{equation} 
is a real-linear subspace of $\hil{H}_{\mu}$.  (Note that $S(G)$ is
not closed nor, as we shall soon see, complex-linear.)  Thus, in the
Newton-Wigner representation, the
classical localization scheme assigns $G$ to the real-linear 
subspace $V(S(G))$ of $L^{2}(\mathbb{R}^{3})$.   When no confusion 
can result, I will suppress reference to the unitary operator $V$ and 
simply use $S(G)$ to denote the pertinent subspace in either concrete
version of the one-particle space.    

Note that the correspondence $G\mapsto S(G)$ is monotone; i.e., if
$G_{1}\subseteq G_{2}$ then $S(G_{1})\subseteq S(G_{2})$.  Moreover,
if $G_{1}\cap G_{2}=\emptyset$, then $S(G_{1})$ and $S(G_{2})$ are
``symplectically orthogonal.''  That is, if $f\in S(G_{1})$ and $g\in
S(G_{2})$, then $\mathrm{Im}(f,g)=0$.  Indeed, if $u_{0}\oplus u_{1}
\in S(G_{1})$ and $v_{0}\oplus v_{1}\in S(G_{2})$, then
\begin{equation} \sigma (u_{0}\oplus u_{1},v_{0}\oplus v_{1})=\int
  _{\mathbb{R}^{3}}(u_{0}v_{1}-u_{1}v_{0})\, d^{3}\mathbf{x} =0 ,\end{equation} since the
$u_{i}$
and $v_{i}$ have disjoint regions of support. 

Now, we say that a Weyl operator $W(f)$ acting on $\hil{F}(\hil{H})$
is \emph{classically localized} in $G$ just in case $f\in S(G)$.
(``Classically'' here refers simply to the fact that our notion of
localization is derived from the local structure of the classical
phase space $S$.)  Let $\mathbf{B}(\hil{F}(\hil{H}))$ denote the
algebra of bounded operators on $\hil{F}(\hil{H})$.  We then define
the subalgebra $\alg{R}(G)\subseteq \mathbf{B}(\hil{F}(\hil{H}))$ of
operators classically localized in $G$ to be the ``von Neumann
algebra'' generated by the Weyl operators classically localized in
$G$.  That is, $\alg{R}(G)$ consists of arbitrary linear combinations
and ``weak limits'' of Weyl operators classically localized in
$G$.\footnote{Since $f\mapsto W(f)$ is weakly continuous, $\alg{R}(G)$
  contains $W(f)$ for all $f$ in the closure of~$S(G)$.}

If $\alg{R}\subseteq \mathbf{B}(\hil{F}(\hil{H}))$, we let
$\alg{R}\,'$ denote all operators in $\mathbf{B}(\hil{F}(\hil{H}))$
that commute with every operator in $\alg{R}$.  If $\alg{R}$ contains
$I$ and is closed under taking adjoints, then von Neumann's ``double
commutant theorem'' entails that $(\alg{R}')'$ is the von Neumann
algebra generated by $\alg{R}$.  Thus, we have
\begin{equation}
\alg{R}(G)= \{ W(f):f\in S(G) \} '' \, . \end{equation}  
In order also to associate unbounded operators with local regions, we say that
an unbounded operator $A$ is \emph{affiliated} with the local algebra
$\alg{R}(G)$ 
just in case $U^{-1}AU=A$ for any unitary operator
$U\in \alg{R}(G)'$.  It then follows that $\Phi (f)$ is affiliated with
$\alg{R}(G)$ just in case $W(f)\in \alg{R}(G)$.  

The correspondence $G\mapsto \alg{R}(G)$ clearly satisfies isotony.
That is, if $G_{1}\subseteq G_{2}$ then $\alg{R}(G_{1})\subseteq
\alg{R}(G_{2})$.  Moreover, the local algebras also satisfy fixed-time
microcausality.  That is, if $G_{1}\cap G_{2}=\emptyset$ then all
operators in $\alg{R}(G_{1})$ commute with all operators in
$\alg{R}(G_{2})$.  (This follows directly from Eq.~(\ref{weyl}) and the
fact that $S(G_{1})$ and $S(G_{2})$ are symplectically orthogonal.)

\subsection{Anti-locality and the Reeh-Schlieder theorem}
Let $\alg{R}$ be some subalgebra of $\mathbf{B}(\hil{F}(\hil{H}))$.
We say that a vector $\psi \in \hil{F}(\hil{H})$ is \emph{cyclic} for
$\alg{R}$ just in case $[\alg{R}\psi ]=\hil{F}(\hil{H})$, where
$[\alg{R}\psi ]$ denotes the closed linear span of $\{ A\psi :A\in
\alg{R} \}$.  Of course, every vector in $\hil{F}(\hil{H})$, including
the vacuum vector $\Omega$, is cyclic for the global algebra
$\mathbf{B}(\hil{F}(\hil{H}))$ of all bounded operators on
$\hil{F}(\hil{H})$.  The Reeh-Schlieder theorem, however, tells us
that the vacuum vector $\Omega$ is cyclic for any \emph{local} algebra
$\alg{R}(G)$.

The first version of the Reeh-Schlieder theorem I will present is a
restricted version of the theorem---due to Segal and
Goodman---applicable only to the free Bose field model.  The key
concept in this version of the theorem is the notion of an
``anti-local'' operator.
\begin{defn} {\it An operator $A$ on $L^{2}(\mathbb{R}^{3})$ is said
    to be anti-local just in case: For any $f\in
    L^{2}(\mathbb{R}^{3})$ and for any open subset $G$ of
    $\mathbb{R}^{3}$, $\mathrm{supp}(f)\cap G=\emptyset$ and
    $\mathrm{supp}(Af)\cap G=\emptyset$ only if $f=0$.} \end{defn}
\noindent Thus, in particular, an anti-local operator maps any wavefunction with support inside a
bounded region to a wavefunction with infinite ``tails.'' 

The following lemma may be the most important lemma for understanding
the local structure of the free Bose field model:

\begin{lemma}[Segal and Goodman 1965] The operator $H=(-\nabla ^{2}+m^{2})^{1/2}$ is
  anti-local.  \end{lemma}

This lemma has the important consequence that for any non-empty open
subset $G$ of $\mathbb{R}^{3}$, the \emph{complex}-linear span of
$S(G)$ is dense in $\hil{H}$ (cf. Segal and Goodman 1965, Corollary
1).  However, for any real-linear subspace $E$ of $\hil{H}$, $\Omega$
is cyclic for the algebra generated by $\{ W(f): f\in E \}$ if and
only if the complex-linear span of $E$ is dense in $\hil{H}$ (cf. Petz
1990, Proposition 7.7).  Thus, the anti-locality of $H$ entails that
$\Omega$ is cyclic for every local algebra.

\begin{rsthm} Let $G$ be any nonempty open subset of
  $\mathbb{R}^{3}$.  Then, $\Omega$ is cyclic for $\alg{R}(G)$.
\end{rsthm}

What is the significance of this cyclicity result?  Segal (1964, 140)
claims that the theorem is ``striking,'' since it entails that
\begin{quote} ``...the entire state vector space of the field could be obtained from measurements
  in an arbitrarily small region of space-time!'' \end{quote} He then
goes on to claim that the result is, ``quite at variance with the
spirit of relativistic causality" (143).  Fleming also sees the
cyclicity result as counterintuitive, apparently because it does not
square well with our understanding of relativistic causality.  For
example (cf. Fleming 2000, 4), the Reeh-Schlieder theorem entails that
for any state $\psi \in \hil{F}(\hil{H})$, and for any predetermined
$\epsilon$, there is an operator $A\in \alg{R}(G)$ such that
$\norm{A\Omega -\psi}<\epsilon$.  In particular, $\psi$ may be a state
that differs from the vacuum only in some region $G\,'$ that is
disjoint (and hence spacelike separated) from $G$.  If, then, $A$ is
interpreted as an ``operation'' that can be performed in the region
$G$, it follows that operations performed in $G$ can result in
arbitrary changes of the state in the region $G\,'$.  This, then, is
taken by Fleming to show that, ``the local fields allow the
possibility of arbitrary space-like distant effects from arbitrary
localized actions'' (Fleming 2000, 20).

Fleming's use of ``actions'' and ``effects'' seems to construe a local
operation---represented by an operator $A\in \alg{R}(G)$---as a purely
\emph{physical} disturbance of the system; i.e., the operation here is
a \emph{cause} with an \emph{effect} at spacelike separation.  If this
were the only way to think of local operations, then I would grant
that the Reeh-Schlieder theorem is counterintuitive, and indeed very
contrary to the spirit of relativisitic causality.  However, once one
makes the crucial distinction between selective and nonselective local
operations, local cyclicity does not obviously conflict with
relativistic causality (cf. Clifton and Halvorson 2000, Section~2).
Rather than dwell on that here, however, I will proceed to spell out
some of the further ``counterintuitive'' consequences of the
Reeh-Schlieder theorem.

\vspace{1em} {\bf 1.}  Let $G_{1}$ and $G_{2}$ be disjoint subsets of
$\mathbb{R}^{3}$. Suppose that $W(f)$ is classically localized in
$G_{1}$ and $W(g)$ is classically localized in $G_{2}$.  Then,
$\mathrm{Im}(f,g)=0$ and therefore $W(f)W(g)=W(f+g)$.  Thus,
\begin{eqnarray} \langle \Omega ,W(f)W(g)\Omega \rangle &=& \exp
  (-\norm{f+g}^{2}/4) \qquad \\
  &=& \langle \Omega ,W(f)\Omega \rangle \cdot \langle \Omega
  ,W(g)\Omega \rangle \cdot e^{-\mathrm{Re} (f,g) /2} \label{entangle}
  .\end{eqnarray} However, $S(G_{1})$ and $S(G_{2})$ are not orthogonal
relative to the real part of the inner product $(\cdot ,\cdot )$.
  Indeed, if $f=u_{0}\oplus u_{1}$
and $g=v_{0}\oplus v_{1}$, then 
\begin{eqnarray}
    \mathrm{Re}(f,g) &=&(u_{0},Hv_{0})+(u_{1},H^{-1}v_{1}) \\
   &=& \int _{\mathbb{R}^{3}} u_{0}(Hv_{0})\,d^{3}\mathbf{x}+\int
   _{\mathbb{R}^{3}}u_{1}(H^{-1}v_{1})\,d^{3}\mathbf{x} . 
\label{integral} \end{eqnarray} 
But since $H$ and $H^{-1}$ are anti-local, the two integrals in
(\ref{integral}) will not generally vanish.  Therefore, the vacuum
state is not a product state across $\alg{R}(G_{1})$ and
$\alg{R}(G_{2})$.

It should be noted, however, that the above argument does not entail
that the vacuum state is ``entangled''---since it could still be a
\emph{mixture} of product states across $\alg{R}(G_{1})$ and
$\alg{R}(G_{2})$.  However, it can be shown directly from the
cyclicity of the vacuum vector $\Omega$ that the vacuum state is not
even a mixture of product states across $\alg{R}(G_{1})$ and
$\alg{R}(G_{2})$ (Halvorson and Clifton 2000).  Moreover, the vacuum
predicts a maximal violation of Bell's inequality relative to the
algebras $\alg{R}(G)$ and $\alg{R}(G\,')$, where
$G\,'=\mathbb{R}^{3}\backslash G$ (Summers and Werner 1985).  (Bell
correlation, however, is not entailed by cyclicity.)  \vspace{1em}

{\bf 2.}  The cyclicity of the vacuum combined with (equal-time)
microcausality entails that the vacuum vector is \emph{separating} for
any local algebra $\alg{R}(G)$, where $G\,'$ has non-empty interior.
That is, for any operator $A\in \alg{R}(G)$, if $A\Omega =0$ then
$A=0$.  In particular, for any local event---represented by projection
operator $P\in \alg{R}(G)$---the probability that event will occur in
the vacuum state is nonzero.  Thus, the vacuum is ``seething with
activity'' at the local level.

Since the vacuum is entangled across $\alg{R}(G)$ and $\alg{R}(G\,')$,
it follows that the vacuum is a mixed state when restricted to the
local algebra $\alg{R}(G)$.  In fact, when we restrict the vacuum to
$\alg{R}(G)$, it is \emph{maximally mixed} in the sense that the
vacuum may be written as a mixture with any one of a dense set of
states of $\alg{R}(G)$ (Clifton and Halvorson 2000).  Intuitively
speaking, then, the vacuum state provides minimal information about
local states of affairs.  This is quite similar to the singlet state,
which restricts to the maximally mixed state $(1/2)I$ on either
one-particle subsystem (cf. Redhead 1995a).  \vspace{1em}
  
{\bf 3.} For any annihilation operator $a(f)$, we have $a(f)\Omega
=0$.  Thus, $a(f)$ cannot be affiliated with the local algebra
$\alg{R}(G)$.  Since the family of operators affiliated with
$\alg{R}(G)$ is closed under taking adjoints, it also follows that no
creation operators are affiliated with $\alg{R}(G)$.

The concreteness of the model we are dealing with allows a more direct
understanding of why, mathematically speaking, local algebras do not
contain creation and annihilation operators.  Inverting the relation
in~(\ref{kraftwerk}), and using the fact that $f\mapsto a^{+}(f)$ is
linear and $f\mapsto a(f)$ is anti-linear, it follows that
\begin{eqnarray} a^{+}(f)&=&2^{-1/2}(\Phi (f)-i\Phi
  (if)) ,\\
  a(f)&=&2^{-1/2}(\Phi (f)+i\Phi (if)) .\end{eqnarray} Thus, an algebra generated by the
operators $\{ W(f):f\in E\}$, will contain the creation and annihilation operators $\{
a^{+}(f),a(f):f\in E\}$ only if $E$ is a \emph{complex}-linear subspace of $\hil{H}$.  This is 
\emph{not} the case for a local algebra $\alg{R}(G)$ where $E=S(G)$
  is a \emph{real}-linear subspace of $\hil{H}$.  In fact, referring
to the concrete one-particle space $\hil{H}_{\mu}$ allows us to see clearly that $S(G)$ is not
invariant under the complex structure $J$.  If $u_{0}\oplus u_{1}\in S(G)$, then 
\begin{equation} J(u_{0}\oplus u_{1})=-H^{-1}u_{1}\oplus Hu_{0} .\end{equation} But since
$H$ and $H^{-1}$ are anti-local, it is not the case that $Hu_{0}\in
C^{\infty}(G)$ or $-H^{-1}u_{1}\in C^{\infty}(G)$.  Thus, $Jf\not\in
S(G)$ when $f\in S(G)$.  What is more, since the complex span of
$S(G)$ is dense in $\hil{H}_{\mu}$, if $S(G)$ \emph{were} a complex
subspace, then it would follow that
$\alg{R}(G)=\mathbf{B}(\hil{F}(\hil{H}))$.

\vspace{1em} {\bf 3.} Number operators also annihilate the vacuum.
Since the vacuum is separating for local algebras, no number operator
is affiliated with any local algebra.  Thus, an observer in the region
$G$ cannot count the number of particles in $G$!

How should we understand the inability of local observers to count the
number of particles in their vicinity?  According to Redhead (1995b), a heuristic calculation shows
that the local number density operator $N_{G}$ does not commute with the
density operator $N_{G\,'}$ (where $G\,'$ is the complement of $G$).
Thus, he claims that
\begin{quote} ``...it is usual in axiomatic formulations of quantum
  field theory to impose a microcausality condition on physically
  significant local observables, \emph{viz} that the associated
  operators \emph{should} commute at space-like separation.  The
  conclusion of this line of argument is that number densities are not
  physical observables, and hence we do not have to bother about
  trying to interpret them.''  (81) \end{quote} 
While Redhead's conclusion is correct, it is instructive to note that his reasoning cannot be
reproduced in a mathematically rigorous fashion.  That is, there are \emph{no} local number
density operators---in particular, neither $N_{G}$ nor $N_{G\,'}$ exist---and so it cannot be
literally true that $N_{G}$ and $N_{G\,'}$ fail to commute.
  
In order to see this, consider first the (single wavefunction) number
operator $N_{f}=a^{+}(f)a(f)$, where $f$ is ``classically localized''
in $G$, i.e., $f\in S(G)$.  Since $f\mapsto a^{+}(f)$ is linear, and
$f\mapsto a(f)$ is anti-linear, it follows that $N_{f}=N_{(e^{it}f)}$
for all $t\in \mathbb{R}$.  That is, a single wavefunction number
operator $N_{f}$ is invariant under phase tranformations of $f$.
However, classical localization of a wavefunction is \emph{not}
invariant under phase transformations.  Thus, it is not possible to
formulate a well-defined notion of classical localization for a single
wavefunction number operator.

How, though, do we define a number density operator $N_{G}$?
Heuristically, one sets \begin{equation} N_{G}=\int _{G}N(\mathbf{x})
  d^{3}\mathbf{x} ,\label{heuristic} \end{equation} where
$N(\mathbf{x})=a^{+}(\mathbf{x})a(\mathbf{x})$.  Since, however,
$N(\mathbf{x})$ is not a well-defined mathematical object,
Eq.~(\ref{heuristic}) is a purely formal expression.  Thus, we replace
$N(\mathbf{x})$ with the single wavefunction number operator $N_{f}$
and we set, \begin{equation} N_{G}=\sum _{i}N_{f_{i}} \, ,
  \label{formal} \end{equation} where $f_{i}$ is a basis of the
real-linear subspace $S(G)$ of $\hil{H}$.\footnote{Actually, this
  infinite sum is also a formal expression, since it sums unbounded
  operators.  A technically correct definition would define $N_{G}$ as
  an upper bound of quadratic forms (see Bratteli and Robinson 1997).}
Using the fact that $N_{f}=N_{if}$ for each $f$, it follows then that
$N_{G}=N_{[G]}$, where $N_{[G]}$ is the number operator for the closed complex-linear span
$[S(G)]$ of $S(G)$ in $\hil{H}$; and the anti-locality of $H$ entails that $[S(G)]=\hil{H}$. 
Therefore, the operator we defined in Eq.~(\ref{formal}) turns out to be the \emph{total} number
operator $N$.  \vspace{1em}
  
{\bf 4.}  The Reeh-Schlieder theorem also has implications for the
\emph{internal} structure of the local algebra $\alg{R}(G)$.  In
particular, the local algebra $\alg{R}(G)$ is what is called a ``type
III'' von Neumann algebra (Araki 1964).  (The algebra
$\mathbf{B}(\hil{F}(\hil{H}))$ of all bounded operators on
$\hil{F}(\hil{H})$ is called a type I von Neumann algebra.)  From a
physical point of view, this is significant since type III algebras
contain only infinite-dimensional projections---which entails that
there are strict limits on our ability to ``isolate'' a local system
from outside influences (Clifton and Halvorson 2000).  Type III
algebras also have \emph{no} pure (normal) states.

\section{Newton-Wigner localization}
In the previous section, we saw that the standard localization scheme
$G\mapsto \alg{R}(G)$ has a number of ``counterintuitive'' features,
all of which follow from the Reeh-Schlieder theorem.  These
counterintuitive features prompted Segal (1964) and Fleming (2000) to
suggest a reworking of the correspondence between spatial regions and
subalgebras of observables.  In this section I give a mathematically
rigorous rendering of the Segal-Fleming proposal, and I show how it
avoids both the Reeh-Schlieder theorem and its consequences.  (Here I
deal only with Fleming's first proposal, prior to his generalization
to ``covariant fields.'')

Recall that a localization scheme defines a correspondence between
regions in space and real-linear subspaces of the one-particle space
$\hil{H}$.  The Newton-Wigner localization scheme defines this
correspondence in precisely the way it is done in elementary quantum
mechanics: A region $G$ in $\mathbb{R}^{3}$ corresponds to the
subspace $L^{2}(G)\subseteq L^{2}(\mathbb{R}^{3})$ of wavefunctions
with probability amplitude vanishing (almost everywhere) outside of
$G$.  We may then use the unitary mapping $V$ between $\hil{H}_{\mu}$
and $L^{2}(\mathbb{R}^{3})$ to identify the subspace $V^{-1}L^{2}(G)$
of Newton-Wigner localized wavefunctions in $\hil{H}_{\mu}$.
Hereafter, I will suppress reference to $V^{-1}$ and use $L^{2}(G)$ to
denote the pertinent subspace in either concrete version of the
one-particle space.

Note that the correspondence $G\mapsto L^{2}(G)$ is monotone; i.e., if
$G_{1}\subseteq G_{2}$ then $L^{2}(G_{1})\subseteq L^{2}(G_{2})$.
Moreover, if $G_{1}\cap G_{2}=\emptyset$, then $L^{2}(G_{1})$ and
$L^{2}(G_{2})$ are \emph{fully} orthogonal---a key difference between
NW localization and classical localization.

Now, we say that a Weyl operator $W(f)$ acting on $\hil{F}(\hil{H})$
is \emph{NW-localized} in $G$ just in case $f \in L^{2}(G)$.  We then
define the algebra $\alg{R}_{NW}(G)$ of NW-localized operators on
$\hil{F}(\hil{H})$ as the von Neumann algebra generated by the Weyl
operators NW-localized in $G$.  That is,
\begin{equation} \alg{R}_{NW}(G)=\{ W(f): f \in L^{2}(G) \} ''\,
  .\end{equation} 
Clearly, the correspondence $G\mapsto \alg{R}_{NW}(G)$ 
satisfies isotony.  Moreover, since $G_{1}\cap G_{2}=\emptyset$ entails that $L^{2}(G_{1})$
and $L^{2}(G_{2})$ are orthogonal subspaces of $\hil{H}$, the correspondence $G\mapsto
\alg{R}_{NW}(G)$ satisfies fixed-time microcausality.  Thus, at
least in this fixed-time formulation, the NW localization scheme appears to have all the
advantages of the classical localization scheme.  I will now proceed to spell 
out some features of the NW localization scheme that may make it 
seem \emph{more} attractive than the standard localization scheme.

If $G$ is an open subset of $\mathbb{R}^{3}$, then \begin{equation}
  L^{2}(\mathbb{R}^{3})=L^{2}(G\cup G\,') =L^{2}(G)\oplus L^{2}(G\,')
  .  \end{equation} Accordingly, if we let
$\hil{F}_{G}=\hil{F}(L^{2}(G))$ and
$\hil{F}_{G\,'}=\hil{F}(L^{2}(G\,'))$ then it follows that
\begin{eqnarray} \hil{F}(\hil{H}) &=& \hil{F}_{G}\otimes
  \hil{F}_{G\,'} . \end{eqnarray} 
(Here the equality sign is intended to denote that there is a natural
isomorphism between $\hil{F}(\hil{H})$ and $\hil{F}_{G}\otimes
\hil{F}_{G\,'}$.)  Moreover, the vacuum vector $\Omega \in \hil{F}(\hil{H})$ is the
product $\Omega _{G}\otimes \Omega _{G\,'}$ of the respective vacuum
vectors in $\hil{F}_{G}$ and $\hil{F}_{G\,'}$.  By definition, $\Phi (f)$ is affiliated with
$\alg{R}_{NW}(G)$ when $f\in L^{2}(G)$.  Since $L^{2}(G)$ is a \emph{complex}-linear
subspace of $\hil{H}$, it follows that $\Phi (if)$ is also affiliated with
$\alg{R}_{NW}(G)$, and hence that $a^{+}(f),a(f),$ and $N_{f}$ are
all affiliated with $\alg{R}_{NW}(G)$.  If we let $U$ denote the unitary
operator that maps $\hil{F}_{G}\otimes \hil{F}_{G\,'}$ naturally onto
$\hil{F}(\hil{H})$, then it is not difficult to see that
\begin{equation} U^{-1}a^{+}(f)U=a_{G}^{+}(f)\otimes I ,\end{equation}
where $a_{G}^{+}(f)$ is the creation operator on $\hil{F}_{G}$.  Thus,
we also have $U^{-1}a(f)U=a_{G}(f)\otimes I$, and since the creation
and annihilation operators $\{ a_{G}^{\pm}(f):f\in L^{2}(G) \}$ form
an irreducible set of operators on $\hil{F}_{G}$, it follows that
\begin{eqnarray} \alg{R}_{NW}(G)&=& \mathbf{B}(\hil{F}_{G})\otimes
  I , \label{type} \\
\alg{R}_{NW}(G\,')&=& I\otimes
  \mathbf{B}(\hil{F}_{G\,'}) .\end{eqnarray}  
(Again, equality here means there is a natural isomorphism.)  

It follows then that acting on $\Omega =\Omega _{G}\otimes \Omega
_{G\,'}$ with elements from $\alg{R}_{NW}(G)$ results only in vectors
of the form $\psi \otimes \Omega _{G\,'}$ for some $\psi \in
\hil{F}_{G}$.  Thus, the vacuum is \emph{not} cyclic for the local
algebra $\alg{R}_{NW}(G)$.

\vspace{1em} {\bf 1.}  It is obvious from the preceding that the
vacuum is a product state across $\alg{R}_{NW}(G)$ and its complement
$\alg{R}_{NW}(G\,')$.  This also follows directly from the fact that
$L^{2}(G)$ and $L^{2}(G\,')$ are fully orthogonal subspaces of
$\hil{H}$.  Indeed, let $W(f)\in \alg{R}_{NW}(G)$ and $W(g)\in
\alg{R}_{NW}(G\,')$.  Then since
$\norm{f+g}^{2}=\norm{f}^{2}+\norm{g}^{2}$, it follows that
\begin{eqnarray} \langle \Omega ,W(f)W(g)\Omega \rangle &=&\langle
  \Omega ,W(f+g) \Omega \rangle \\
&=& \exp ( -\norm{f+g}^{2}/4 )  \\
&=& \langle \Omega ,W(f)\Omega \rangle \cdot \langle \Omega ,W(g)
\Omega \rangle . \end{eqnarray}

\vspace{1em} {\bf 2.}  Restricting the vacuum state $\Omega$ to
$\alg{R}_{NW}(G)$ is equivalent to restricting the product state
$\Omega _{G}\otimes \Omega _{G\,'}$ to $\mathbf{B}(\hil{F}_{G})\otimes
I$.  Thus, the restriction of $\Omega$ to $\alg{R}_{NW}(G)$ is pure,
and the global vacuum provides a ``maximally specific'' description of
local states of affairs.

\vspace{1em} {\bf 3.}  If $\{ f_{i}\}$ is an orthonormal basis of
$L^{2}(G)$, then the number operator $N_{G}=\sum _{i}N_{f_{i}}$ is
affiliated with $\alg{R}_{NW}(G)$.  Moreover, the number operator
$N_{G\,'}$ is affiliated with $\alg{R}_{NW}(G\,')$, and by
microcausality we have $[N_{G},N_{G\,'}]=0$.  We may also see this by
employing the correspondence between $\hil{F}(\hil{H})$ and
$\hil{F}_{G}\otimes \hil{F}_{G\,'}$.  The Fock space $\hil{F}_{G}$ has
its own total number operator $\widetilde{N}_{G}$.  Similarly,
$\hil{F}_{G\,'}$ has its own total number operator
$\widetilde{N}_{G\,'}$.  Obviously then, $\widetilde{N}_{G}\otimes I$
is affiliated with $\mathbf{B}(\hil{F}_{G})\otimes I$, and $I\otimes
\widetilde{N}_{G\,'}$ is affiliated with $I\otimes
\mathbf{B}(\hil{F}_{G\,'})$.  Just as obviously,
$\widetilde{N}_{G}\otimes I$ commutes with $I\otimes
\widetilde{N}_{G\,'}$.

\vspace{1em} {\bf 4.}  As can be seen from Eq.~(\ref{type}), the local
algebra $\alg{R}_{NW}(G)$ is a type I von Neumann algebra.  According
to Segal (1964, 140), this is precisely the structure of local
algebras that is ``suggested by considerations of causality and
empirical accessibility.''

\section{The full strength of Reeh-Schlieder}
The results of the previous two sections speak for themselves: The
Newton-Wigner localization scheme results in a mathematical structure
that appears to be much more in accord with our a priori physical
intuitions than the structure obtained from the standard localization
scheme.  In this section, however, I show that the NW localization
scheme ``avoids'' the Reeh-Schlieder theorem in only a trivial sense,
and I show that the NW localization scheme has its own
counterintuitive features without parallel in the standard
localization scheme.

First, while the NW-local algebras avoid cyclicity of the vacuum
vector, they still have a dense set of cyclic vectors.\footnote{Cf.
  Fleming's claim that, ``...it is remarkable that \emph{any state}
  can have enough structure within an arbitrarily small region, $O$,
  to enable even the mathematical reconstituting of essentially the
  whole state space'' (Fleming 2000, 5).}
\begin{thm} $\alg{R}_{NW}(G)$ has a dense set of cyclic vectors in
  $\hil{F}(\hil{H})$.  \label{dense} \end{thm}
\begin{proof} Since the Hilbert spaces $\hil{F}_{G}$ and
  $\hil{F}_{G\,'}$ have the same (infinite) dimension, it follows from
  Theorem 4 of (Clifton et al. 1998) that
  $\alg{R}_{NW}(G)=\mathbf{B}(\hil{F}_{G})\otimes I$ has a dense set
  of cyclic vectors in $\hil{F}(\hil{H})=\hil{F}_{G}\otimes
  \hil{F}_{G\,'}$.  \end{proof} Thus, if the worry about the
Reeh-Schlieder theorem is about cyclicity in general, adopting the NW
localization scheme does nothing to alleviate this worry.

Perhaps, however, the worry about the Reeh-Schlieder theorem is
specifically a worry about cyclicity of the \emph{vacuum} state.  (One
wonders, though, why this would be worse than cyclicity of any other
state.)  Even so, I argue now that the NW localization scheme does not
avoid the ``vacuum-specific'' consequences of the full Reeh-Schlieder
theorem.

Let $\hil{K}$ be an arbitrary Hilbert space, representing the state
space of some quantum field theory.  (For example,
$\hil{K}=\hil{F}(\hil{H})$ in the case of the free Bose field.)
Suppose also that there is a representation $\mathbf{a}\mapsto
U(\mathbf{a})$ of the spacetime translation group in the group of
unitary operators on $\hil{K}$.  Given such a representation, there is
a ``four operator'' $\mathbf{P}$ on $\hil{K}$ such that
$U(\mathbf{a})=e^{i\mathbf{a}\cdot \mathbf{P}}$.  We say that the
representation $\mathbf{a}\mapsto U(\mathbf{a})$ satisfies the
\emph{spectrum condition} just in case the spectrum of $\mathbf{P}$ is
contained in the forward light cone.  From a physical point of view,
the spectrum condition corresponds to the assumption that (a) all
physical effects propagate at velocities at most the speed of light,
and (b) energy is positive.  Note, consequently, that the spectrum
condition is a purely global condition, and so is not likely to be a
source of dispute between proponents of differing localization
schemes.

A \emph{net of local observable algebras} is an assigment $O\mapsto
\alg{A}(O)$ of open regions in Minkowski spacetime to von Neumann
subalgebras of $\mathbf{B}(\hil{K})$.  (Note that this definition is
not immediately pertinent to the localization schemes presented in
Sections~3 and~4, since they gave an assignment of algebras to open
regions in space at a fixed time.)  The full Reeh-Schlieder theorem
will apply to this net if it satisfies the following postulates:
\begin{enumerate}
\item \emph{Isotony:} If $O_{1}\subseteq O_{2}$, then
  $\alg{A}(O_{1})\subseteq \alg{A}(O_{2})$. 
\item \emph{Translation Covariance:}
  $U(\mathbf{a})^{-1}\alg{A}(O)U(\mathbf{a})=\alg{A}(O+\mathbf{a})$.
\item \emph{Weak Additivity:} For any open $O\subseteq M$, the set \[
  \bigcup _{\mathbf{a}\in M}U(\mathbf{a})^{-1}\alg{A}(O)U(\mathbf{a})
  \] of operators is irreducible (i.e., leaves no subspace of
  $\hil{K}$ invariant).
\end{enumerate}
In this general setting, a vacuum vector $\Omega$ can be taken to be
any vector invariant under all spacetime translations
$U(\mathbf{a})$. 
\begin{fullrs} Suppose that $\{ \alg{A}(O)\}$ is a net of local 
  observable algebras satisfying postulates 1--3.  Then, for any open region
  $O$ in Minkowski spacetime, $\Omega$ is cyclic for $\alg{A}(O)$.
\end{fullrs}
\noindent Note that the Reeh-Schlieder theorem does \emph{not} require the postulate of
microcausality (i.e., if $A\in \alg{A}(O_{1})$ and $B\in \alg{A}(O_{2})$, where $O_{1}$ and
$O_{2}$ are spacelike separated, then $[A,B]=0$).\footnote{To see that microcausality is
logically independent from postulates 1--3, take the trivial localization scheme:
$\alg{A}(O)=\mathbf{B}(\hil{K})$, for each $O$.}

For the standard localization scheme, there is a straightforward
connection between the full Reeh-Schlieder theorem and the fixed-time
version given in Section~3.  In particular, there is an alternative
method for describing the standard localization scheme that involves
appeal to spacetime regions rather than space regions at a fixed time
(see Horuzhy 1988, Chapter 4).  It then follows that
$\alg{R}(G)=\alg{A}(O_{G})$, where $O_{G}$ is the ``domain of
dependence'' of the spatial region $G$.  Thus, the fixed-time version
of the Reeh-Schlieder theorem may be thought of as corollary of the
full Reeh-Schlieder theorem in connection with the fact that
$\alg{R}(G)=\alg{A}(O_{G})$.

Segal and Fleming avoid the fully general version of the
Reeh-Schlieder theorem only by remaining silent about how we ought to
assign algebras of observables to open regions of
\emph{spacetime}.\footnote{It is essential for the proof of the full
  Reeh-Schlieder theorem that the region $O$ has some ``temporal
  extension'': The theorem uses the fact that if $A\in \alg{A}(O_{1})$
  where $O_{1}\subset O$, then $U(\mathbf{a})^{-1}AU(\mathbf{a})\in
  \alg{A}(O)$ for sufficiently small $\mathbf{a}$ in four independent
  directions.}  Since, however, the typical quantum field theory
cannot be expected to admit a fixed-time ($3+1$) formulation (cf. Haag
1992, 59), it is not at all clear that they have truly avoided the
Reeh-Schlieder theorem in any interesting sense.  It would certainly
be interesting to see which, \emph{if any}, of the full Reeh-Schlieder
theorem's three premises would be rejected by a more general NW
localization scheme.

However, we need not speculate about the possibility that the full
Reeh-Schlieder theorem will apply to some generalization of NW
localization scheme: The Reeh-Schlieder theorem already has
counterintuitive consequences for the fixed-time NW localization
scheme.  In particular, although the vacuum $\Omega$ is not cyclic
under operations NW-localized in some spatial region $G$ at a single
time, $\Omega$ \emph{is} cyclic under operators NW-localized in $G$
\emph{within an arbitrary short time interval}.  Before I give the
precise version of this result, I should clarify some matters
concerning the relationship between the dynamics of the field and
local algebras.

In the standard localization scheme, the dynamics of local algebras
may be thought of two ways.  On the one hand, we may think of the
assignment $G\mapsto \alg{R}(G)$ as telling us, once and for all, which
observables are associated with the region $G$, in which case the state of $\alg{R}(G)$ (i.e., the
reduced state of the entire field) changes via the unitary evolution $U(t)$ (Schr{\"o}dinger
picture).  On the other hand, we may think of the state of the field as fixed, in which case the
algebra $\alg{R}(G)$ evolves over time to the algebra $U(t)^{-1}\alg{R}(G)U(t)$ (Heisenberg
picture).  Thus, $U(t)^{-1}\alg{R}(G)U(t)$ gives those operators classically localized in $G$ at
time $t$.  The Schr{\"o}dinger picture is particularly intuitive in this case, since it mimics the
dynamics of a classical field where quantities associated with points in space change their values
over time.  

Now, neither Segal nor Fleming explain how we should think of the
dynamics of the NW-local algebras.  Presumably, however, we are to
think of the dynamics of the NW-local algebras in precisely the same
way as we think of the dynamics of the standard local
algebras.\footnote{It is conceivable that Segal or Fleming have some
  different idea concerning the relationship between NW-local algebras
  at different times.  For example, perhaps even in the
  Schr{\"o}dinger picture, the map $G\mapsto \alg{R}_{NW}(G)$ should
  be thought of as time-dependent.  Although this is surely a formal
  possibility, it is exceedingly difficult to understand what it might
  mean, physically, to have a time-dependent association of physical magnitudes with regions in
space.}  In particular, we may
suppose that the state of the field is, at all times, the vacuum state
$\Omega$, and that $U(t)^{-1}\alg{R}_{NW}(G)U(t)$ gives those
operators NW-localized in $G$ at time $t$.

Now for any $\Delta \subseteq \mathbb{R}$ let
\begin{equation} \mathbf{S}_{\Delta }=\{ U(t)^{-1}AU(t): A\in
  \alg{R}_{NW}(G), \,t\in \Delta \} \,.
\end{equation} That is, $\mathbf{S}_{\Delta }$ consists of those operators NW-localized in $G$
at some time $t\in \Delta $.
\begin{thm} For any interval $(a,b)$ around
  $0$, $\Omega$ is cyclic for $\mathbf{S}_{(a,b)}$.  \end{thm}

\begin{proof}[Sketch of proof:]  Let $[\mathbf{S}_{(a,b)}\Omega ]$ denote the closed
  linear span of $\{ A\Omega :A\in \mathbf{S}_{(a,b)} \}$.  Since the
  infinitesimal generator $d\Gamma (H)$ of the group $U(t)$ is
  positive, Kadison's ``little Reeh-Schlieder theorem'' (1970) entails
  that $[\mathbf{S}_{(a,b)}\Omega ]=[\mathbf{S}_{\mathbb{R}}\Omega ]$.
  However, $[\mathbf{S}_{\mathbb{R}}\Omega ]=\hil{F}(\hil{H})$; i.e.,
  $\Omega$ is cyclic under operators NW-localized in $G$ over all
  times (Segal 1964, 143).  Therefore, $\Omega$ is cyclic for
  $\mathbf{S}_{(a,b)}$.
\end{proof}

In Fleming's language, then, the NW-local fields ``allow the
possibility of arbitrary space-like distant effects'' from actions
localized in an arbitrarily small region of space over an arbitrarily
short period of time.  Is this any less ``counterintuitive'' than the
instantaneous version of the Reeh-Schlieder theorem for the standard
localization scheme?\footnote{One may, however, reject the
  interpretation of elements of $\alg{R}_{NW}(G)$ as operations that
  can be performed in $G$.  I return to this point in the next
  section.}
 
Finally, we are in a position to see explicitly a ``counterintuitive''
feature of the NW localization scheme that is not shared by the
standard localization scheme: NW-local operators fail to commute at
spacelike separation.  For this, choose mutually disjoint regions
$G_{1}$ and $G_{2}$ in $\mathbb{R}^{3}$, and choose an interval
$(a,b)$ around $0$ so that $O_{1}:=\cup _{t\in (a,b)}(G_{1}+t)$ and
$O_{2}:=\cup _{t\in (a,b)}(G_{2}+t)$ are spacelike separated.  Let
$\alg{A}_{NW}(O_{i})$ be the von Neumann algebra generated by
  \begin{equation}
\bigcup _{t\in (a,b)} U(t)^{-1}\alg{R}_{NW}(G_{i})U(t) .\end{equation} 
Then it follows from Theorem~2 that the vacuum is cyclic for
$\alg{A}_{NW}(O_{2})$.  However, since $\alg{A}_{NW}(O_{1})\supseteq
\alg{R}_{NW}(G)$ contains annihilation operators and number operators,
it follows that $\alg{A}_{NW}(O_{1})$ and $\alg{A}_{NW}(O_{2})$ do not
satisfy microcausality.  (Microcausality, in conjunction with cyclicity of
the vacuum vector, would entail that the vacuum vector is separating.)
More specifically, while the algebras $U(t)^{-1}\alg{R}_{NW}(G_{1})U(t)$ and
$U(t)^{-1}\alg{R}_{NW}(G_{2})U(t)$ do satisfy microcausality for any
fixed $t$,  microcausality does not generally hold for the algebras
$U(t)^{-1}\alg{R}_{NW}(G_{1})U(t)$ and
$U(s)^{-1}\alg{R}_{NW}(G_{2})U(s)$ when $t\neq s$ (despite the fact
that $G_{1}+t$ and $G_{2}+s$ are spacelike separated).

It would be naive at this stage to claim that failure of generalized microcausality provides a simple
reductio on the NW-localization scheme.  As I will argue in the
next section, however, the failure of generalized microcausality for
the NW-local algebras leaves little room for making any physical sense
of the NW localization scheme.

\section{Local properties and local measurements}
The assignment $G\mapsto \alg{R}(G)$ was originally taken to have the
operational meaning that $\alg{R}(G)$ consists of those observables
that are \emph{measurable} in the region $G$.  What is the intended
meaning of the alternative assignment $G\mapsto \alg{R}_{NW}(G)$?
Segal (1964, 142) presents the NW localization scheme as a contrasting
claim about what can be measured in the spatial region $G$:
\begin{quote} ``From an operational viewpoint it is these variables
  [i.e., $\Phi (f)$ with $f\in L^{2}(G)$]...that appear as the
  localized field variables, and the ring $\alg{R}_{NW}(G)$...appears
  as the appropriate ring of local field observables, rather than the
  ring $\alg{R}(G)$...''  (notation adapted)
\end{quote}
According to this interpretation, the standard localization scheme and
the NW localization scheme present us with two empirically
\emph{in}equivalent versions of quantum field theory.  (For example,
the vacuum displays Bell correlations relative to the algebras
$\alg{R}(G)$ and $\alg{R}(G\,')$, while the vacuum is a product state
across $\alg{R}_{NW}(G)$ and $\alg{R}_{NW}(G\,')$.)  Thus, deciding
which localization scheme is ``correct'' would be a matter of
experiment, not a matter of interpretation.

There is, however, also a conceptual difficulty with interpreting
$\alg{R}_{NW}(G)$ as the algebra of observables measurable in $G$.  In
particular, if $[A,B]\neq 0$, then measurement of $A$ can affect the
statistics for outcomes of $B$ and vice versa.  Thus, if
$\alg{R}_{NW}(G)$ is what is measurable in $G$, then the failure of
generalized microcausality for the NW-local algebras would pose the
threat of causal anomalies.  

This difficulty posed by the failure of generalized microcausality is
clear to Fleming.  In response, he and Butterfield (1999, 158) note
that
\begin{quote} ``...one naturally assumes that one can interpret the  
  \emph{association} of an operator with a spacetime region as
  implying that one can \emph{measure it by performing operations
    confined} to that region,'' 
\end{quote} and they assert that they, ``...question [this] interpretive 
assumption'' (159).  But if not local measurability, what does
association of an observable with a spatial region mean?  Fleming
(2000, 21) claims that NW position operators,
\begin{quote} ``...are more closely related than the local field coordinate to assessments
  of \emph{where}, on hyperplanes and in space-time, objects, systems,
  their localizable properties and phenomena are located.''
\end{quote} It seems then that Fleming intends something along the lines of:
\begin{quote}
  $(\dagger )$ The projections in $\alg{R}_{NW}(G)$ correspond to the
  \emph{properties} of the system that are \emph{localized in $G$.}
\end{quote} But what does Fleming mean by saying that a property is localized in a
spatial region $G$?  And why would the properties localized in $G$
differ from what can be measured in $G$?

Although Fleming has not offered a ``philosophical account'' of
localized properties, he has provided analogies from classical
mechanics in order to prime our intuitions about physical quantities
that may ``pertain to'' a region, without being measurable in that
region (cf. Fleming 2000; Fleming and Butterfield 1999).  For example,
take the center of mass $C$ of a spatially extended system.  At a
given time, $C$ is located at a point $\mathbf{x}$ in space, but $C$
is not measurable at $\mathbf{x}$ or even in spatial regions
immediately surrounding $\mathbf{x}$.  Perhaps then we can think of
NW-localized quantities as similar to center of mass, center of
charge, and their ilk.

This analogy, however, conceals an equivocation in the meaning of
``localized.''  To see this clearly, let me distinguish two types of
localized quantities in classical mechanics.  On the one hand, a
physical quantity $Q$ may be permanently attached to some point
$\mathbf{x}$ in space, in which case the values of $Q$ are given by
vectors (or more generally, tensors) in the tangent space
$T_{\mathbf{x}}$ over $\mathbf{x}$ (e.g., magnetic field strength at
$\mathbf{x}$).  I will refer to this first type of localized quantity as
\emph{\underline{f}ixedly-localized}.  On the other hand, some
quantities take vectors in \emph{physical space} as their values
(e.g., center of mass of a spatially extended system, or position of a
point particle).  I will refer to this latter type of localized quantity
as \emph{\underline{v}ariably-localized}.

I grant that there is a sense in which v-localized quantities are
``localized,'' despite the fact that they may not be locally
measurable (e.g., center of mass).  However, analogies to v-localized
quantities go no distance in clarifying the localization map $G\mapsto
\alg{R}_{NW}(G)$, since this is a permanent assignment
(f-localization) of physical quantities to regions in space.  What we
really need, then, is an example of a f-localized quantity that is not
locally measurable.

One might claim that examples of such quantities are readily
forthcoming:  Let $C$ be center of mass, and let $Q$ be the quantity
that assumes the value $1$ if $C=\mathbf{x}$, and $0$ if $C\neq
\mathbf{x}$.  Obviously, $Q$ will not typically be measurable in the
vicinity of $\mathbf{x}$.  But shouldn't we say that $Q$ always
``pertains to'' $\mathbf{x}$, or is f-localized at $\mathbf{x}$?

The intuition behind thinking that $Q$ ``pertains to'' $\mathbf{x}$
seems to be based on the fact that $Q$ tells us something about
$\mathbf{x}$, viz., whether it is $C$'s value.  However, if this a
sufficient condition for $Q$'s pertaining to $\mathbf{x}$, then $Q$
pertains to \emph{every} point in space.  Indeed, let $\mathbf{y}$ be
another vector in $\mathbb{R}^{3}$ and introduce the new quantity
$C\,'=C+(\mathbf{y}-\mathbf{x})$.  Then $Q=1$ if and only if
$C\,'=\mathbf{y}$, and so the previous line of reasoning would imply
that $Q$ pertains to $\mathbf{y}$ (since $Q$ tells us whether
$\mathbf{y}$ is the value of $C\,'$).  What we should conclude, then,
is that $Q$ is \emph{not} f-localized at $\mathbf{x}$ in the same
sense that a field quantity may be localized at $\mathbf{x}$.  Thus,
we have yet to find an example of a f-localized quantity that is not
locally measurable.

In summary, while it is clear what it means for a physical quantity to
be v-localized in $G$ but not measurable in $G$, it is by no means
clear what it would mean for a physical quantity to be f-localized in
$G$ but not measurable in $G$.  As a result, it is also not clear how we
should interpret the localization map $G\mapsto \alg{R}_{NW}(G)$.

\section{Conclusion}
Introduction of the NW localization scheme into quantum field theory
was an ingenious move.  By means of one deft transformation, it
appears to thwart the Reeh-Schlieder theorem and to restore the
``intuitive'' picture of localization from non-relativistic quantum
mechanics.  However, there are many reasons to doubt that
Newton-Wigner has truly spared us of the counterintuitive consequences
of the Reeh-Schlieder theorem.  First, NW-local algebras still have a
dense set of cyclic vectors.  Second, since general quantum field
theories cannnot be expected to admit a fixed-time formulation, it is
not clear that the NW localization scheme has any interesting level of
generality.  Third, NW-local operations on the vacuum over an
arbitrarily short period of time do generate the state space of the
entire field.  And, finally, the failure of generalized microcausality
for the NW localization scheme leaves us without any natural physical
interpretation of the correspondence $G\mapsto \alg{R}_{NW}(G)$.

After showing that the Reeh-Schlieder theorem fails for NW-local
algebras, Fleming (2000, 11) states that, ``Now it is clear why it
would be worthwhile to see the NW fields as covariant structures.''
While there may be very good reasons for seeing the NW fields as
covariant structures, avoiding the Reeh-Schlieder theorem is
\emph{not} one of them.  \vspace{1em}

{\it Acknowledgments:} I am grateful for helpful correspondence from Jeremy Butterfield and
Bernard Kay, and I am extremely grateful to Rob Clifton for his input throughout this project.   
\vspace{1em}
 
\noindent \textbf{References:}

Araki, Huzihiro (1964), ``Types of von Neumann algebras of local observables
for the free scalar field'', {\it Progress of Theoretical Physics} 32:
956-961.

Bratteli, Ola, and Derek Robinson (1997), {\it Operator Algebras and
  Quantum Statistical Mechanics}, Vol. 2.  NY: Springer.

Clifton, Rob, David Feldman, Hans Halvorson, Michael Redhead, and Alex
Wilce (1998), ``Superentangled states'', {\it Physical Review A} 58:
135-145.

Clifton, Rob, and Hans Halvorson (2000), ``Entanglement and open systems
in algebraic quantum field theory'', {\it Studies in the History and
  Philosophy of Modern Physics}, forthcoming.

Fleming, Gordon (1996), ``Just how radical is hyperplane dependence?'', in
Rob Clifton (ed.), {\it Perspectives on Quantum Reality.} Dordrecht:
Kluwer.

Fleming, Gordon (2000), ``Reeh-Schlieder meets Newton-Wigner'', in
{\it PSA 1998}, v. 2, forthcoming.

Fleming, Gordon, and Jeremy Butterfield (1999), ``Strange positions'', in Jeremy Butterfield and
Constantine Pagonis (eds.), {\it From Physics to Philosophy},
NY: Cambridge University Press.

Haag, Rudolf (1992), {\it Local Quantum Physics.} NY: Springer.

Halvorson, Hans, and Rob Clifton (2000), ``Generic Bell correlation
between arbitrary local algebras in quantum field theory'', {\it
  Journal of Mathematical Physics} 41: 1711-1717.

Horuzhy, Sergei (1988), {\it Introduction to Algebraic Quantum Field
  Theory.}  Dordrecht: Kluwer.

Kadison, Richard V. (1970), ``Some analytic methods in the theory of
operator algebras'', in C. T. Taam (ed.), {\it Lectures in Modern
  Analysis and Applications}, Vol. II.  NY: Springer, 8-29.

Kay, Bernard (1979), ``A uniqueness result in the Segal-Weinless approach
to linear Bose fields'', {\it Journal of Mathematical Physics} 20:
1712-1713.

Malament, David (1996), ``In defense of dogma: Why there cannot be a
relativistic quantum mechanics of (localizable) particles'', in Rob Clifton (ed.), {\it Perspectives on
Quantum Reality.} Dordrecht:
Kluwer, 1-10.

Petz, D{\'e}nes (1990), {\it An Invitation to the Algebra of Canonical
  Commutation Relations}.  Leuven University Press.

Redhead, Michael (1995a), ``More ado about nothing'', {\it Foundations
  of Physics} 25: 123-137.

Redhead, Michael (1995b), ``The vacuum in relativistic quantum field
theory'', in David Hull, Micky Forbes, and Richard M. Burian (eds.),
{\it PSA 1994}, v. 2. East Lansing, MI: Philosophy of Science
Association, 77-87.

Saunders, Simon (1992), ``Locality, complex numbers, and relativistic
quantum theory'', in David Hull, Micky Forbes, and Kathleen Okruhlik (eds.), {\it PSA 1992}, v.
1: East Lansing,
MI: Philosophy of Science Association, 365-380.

Segal, Irving E. (1964), ``Quantum fields and analysis in the solution
manifolds of differential equations'', in William T. Martin and Irving E. Segal,
(eds.), {\it Proceedings of a Conference on the Theory and Applications
  of Analysis in Function Space}.  Cambridge: MIT Press, 129-153.

Segal, Irving E., and Roe W. Goodman (1965), ``Anti-locality of certain
Lorentz-invariant operators'', {\it Journal of Mathematics and
  Mechanics} 14: 629-638.

Summers, Stephen J., and Reinhard Werner (1985), ``The vacuum violates
Bell's inequalities'', {\it Physics Letters} 110A: 257-259.

\end{document}

%% file: diagram.tex
\makeatletter
 
\def\diagram{\m@th\leftwidth=\z@ \rightwidth=\z@ \topheight=\z@
\botheight=\z@ \setbox\@picbox\hbox\bgroup}
 
\def\enddiagram{\egroup\wd\@picbox\rightwidth\unitlength
\ht\@picbox\topheight\unitlength \dp\@picbox\botheight\unitlength
\hskip\leftwidth\unitlength\box\@picbox}
 
\def\bfig{\begin{diagram}}
\def\efig{\end{diagram}}
\newcount\wideness \newcount\leftwidth \newcount\rightwidth
\newcount\highness \newcount\topheight \newcount\botheight
 
\def\ratchet#1#2{\ifnum#1<#2 \global #1=#2 \fi}
 
\def\putbox(#1,#2)#3{%
\horsize{\wideness}{#3} \divide\wideness by 2
{\advance\wideness by #1 \ratchet{\rightwidth}{\wideness}}
{\advance\wideness by -#1 \ratchet{\leftwidth}{\wideness}}
\vertsize{\highness}{#3} \divide\highness by 2
{\advance\highness by #2 \ratchet{\topheight}{\highness}}
{\advance\highness by -#2 \ratchet{\botheight}{\highness}}
\put(#1,#2){\makebox(0,0){$#3$}}}
 
\def\putlbox(#1,#2)#3{%
\horsize{\wideness}{#3}
{\advance\wideness by #1 \ratchet{\rightwidth}{\wideness}}
{\ratchet{\leftwidth}{-#1}}
\vertsize{\highness}{#3} \divide\highness by 2
{\advance\highness by #2 \ratchet{\topheight}{\highness}}
{\advance\highness by -#2 \ratchet{\botheight}{\highness}}
\put(#1,#2){\makebox(0,0)[l]{$#3$}}}
 
\def\putrbox(#1,#2)#3{%
\horsize{\wideness}{#3}
{\ratchet{\rightwidth}{#1}}
{\advance\wideness by -#1 \ratchet{\leftwidth}{\wideness}}
\vertsize{\highness}{#3} \divide\highness by 2
{\advance\highness by #2 \ratchet{\topheight}{\highness}}
{\advance\highness by -#2 \ratchet{\botheight}{\highness}}
\put(#1,#2){\makebox(0,0)[r]{$#3$}}}

\def\adjust[#1]{} % For compatibility
 
\newcount \coefa
\newcount \coefb
\newcount \coefc
\newcount\tempcounta
\newcount\tempcountb
\newcount\tempcountc
\newcount\tempcountd
\newcount\xext
\newcount\yext
\newcount\xoff
\newcount\yoff
\newcount\gap%
\newcount\arrowtypea
\newcount\arrowtypeb
\newcount\arrowtypec
\newcount\arrowtyped
\newcount\arrowtypee
\newcount\height
\newcount\width
\newcount\xpos
\newcount\ypos
\newcount\run
\newcount\rise
\newcount\arrowlength
\newcount\halflength
\newcount\arrowtype
\newdimen\tempdimen
\newdimen\xlen
\newdimen\ylen
\newsavebox{\tempboxa}%
\newsavebox{\tempboxb}%
\newsavebox{\tempboxc}%
 
\newdimen\w@dth
 
\def\setw@dth#1#2{\setbox\z@\hbox{\m@th$#1$}\w@dth=\wd\z@
\setbox\@ne\hbox{\m@th$#2$}\ifnum\w@dth<\wd\@ne \w@dth=\wd\@ne \fi
\advance\w@dth by 1.2em}
 
\def\t@^#1_#2{\allowbreak\def\n@one{#1}\def\n@two{#2}\mathrel
{\setw@dth{#1}{#2}
\mathop{\hbox to \w@dth{\rightarrowfill}}\limits
\ifx\n@one\empty\else ^{\box\z@}\fi
\ifx\n@two\empty\else _{\box\@ne}\fi}}
\def\t@@^#1{\@ifnextchar_{\t@^{#1}}{\t@^{#1}_{}}}
\def\to{\@ifnextchar^{\t@@}{\t@@^{}}}
 
\def\t@left^#1_#2{\def\n@one{#1}\def\n@two{#2}\mathrel{\setw@dth{#1}{#2}
\mathop{\hbox to \w@dth{\leftarrowfill}}\limits
\ifx\n@one\empty\else ^{\box\z@}\fi
\ifx\n@two\empty\else _{\box\@ne}\fi}}
\def\t@@left^#1{\@ifnextchar_{\t@left^{#1}}{\t@left^{#1}_{}}}
\def\toleft{\@ifnextchar^{\t@@left}{\t@@left^{}}}
 
\def\two@^#1_#2{\allowbreak
\def\n@one{#1}\def\n@two{#2}\mathrel{\setw@dth{#1}{#2}
\mathop{\vcenter{\lineskip\z@\baselineskip\z@
                 \hbox to \w@dth{\rightarrowfill}%
                 \hbox to \w@dth{\rightarrowfill}}%
       }\limits
\ifx\n@one\empty\else ^{\box\z@}\fi
\ifx\n@two\empty\else _{\box\@ne}\fi}}
\def\tw@@^#1{\@ifnextchar _{\two@^{#1}}{\two@^{#1}_{}}}
\def\two{\@ifnextchar ^{\tw@@}{\tw@@^{}}}
 
\def\tofr@^#1_#2{\def\n@one{#1}\def\n@two{#2}\mathrel{\setw@dth{#1}{#2}
\mathop{\vcenter{\hbox to \w@dth{\rightarrowfill}\kern-1.7ex
                 \hbox to \w@dth{\leftarrowfill}}%
       }\limits
\ifx\n@one\empty\else ^{\box\z@}\fi
\ifx\n@two\empty\else _{\box\@ne}\fi}}
\def\t@fr@^#1{\@ifnextchar_ {\tofr@^{#1}}{\tofr@^{#1}_{}}}
\def\tofro{\@ifnextchar^ {\t@fr@}{\t@fr@^{}}}

\def\mon{\mathop{\m@th\hbox to
      14.6\P@{\lasyb\char'51\hskip-2.1\P@$\arrext$\hss
$\mathord\rightarrow$}}\limits} % width of \epi
\def\leftmono{\mathrel{\m@th\hbox to
14.6\P@{$\mathord\leftarrow$\hss$\arrext$\hskip-2.1\P@\lasyb\char'50%
}}\limits} % width of \epi
\mathchardef\arrext="0200       % amr minus for arrow extension (see \into)

\def\settypes(#1,#2,#3){\arrowtypea#1 \arrowtypeb#2 \arrowtypec#3}
\def\settoheight#1#2{\setbox\@tempboxa\hbox{#2}#1\ht\@tempboxa\relax}%
\def\settodepth#1#2{\setbox\@tempboxa\hbox{#2}#1\dp\@tempboxa\relax}%
\def\settokens`#1`#2`#3`#4`{%
     \def\tokena{#1}\def\tokenb{#2}\def\tokenc{#3}\def\tokend{#4}}
\def\setsqparms[#1`#2`#3`#4;#5`#6]{%
\arrowtypea #1
\arrowtypeb #2
\arrowtypec #3
\arrowtyped #4
\width #5
\height #6
}
\def\setpos(#1,#2){\xpos=#1 \ypos#2}

\def\settriparms[#1`#2`#3;#4]{\settripairparms[#1`#2`#3`1`1;#4]}%
 
\def\settripairparms[#1`#2`#3`#4`#5;#6]{%
\arrowtypea #1
\arrowtypeb #2
\arrowtypec #3
\arrowtyped #4
\arrowtypee #5
\width #6
\height #6
}
 
\def\resetparms{\settripairparms[1`1`1`1`1;500]\width 500}%default values%
 
\resetparms
 
\def\mvector(#1,#2)#3{%%
\put(0,0){\vector(#1,#2){#3}}%
\put(0,0){\vector(#1,#2){26}}%
}
\def\evector(#1,#2)#3{{%%
\arrowlength #3
\put(0,0){\vector(#1,#2){\arrowlength}}%
\advance \arrowlength by-30
\put(0,0){\vector(#1,#2){\arrowlength}}%
}}
 
\def\horsize#1#2{%
\settowidth{\tempdimen}{$#2$}%
#1=\tempdimen
\divide #1 by\unitlength
}
 
\def\vertsize#1#2{%
\settoheight{\tempdimen}{$#2$}%
#1=\tempdimen
\settodepth{\tempdimen}{$#2$}%
\advance #1 by\tempdimen
\divide #1 by\unitlength
}
 
\def\putvector(#1,#2)(#3,#4)#5#6{{%
\ifnum3<\arrowtype
\putdashvector(#1,#2)(#3,#4)#5\arrowtype
\else
\ifnum\arrowtype<-3
\putdashvector(#1,#2)(#3,#4)#5\arrowtype
\else
\xpos=#1
\ypos=#2
\run=#3
\rise=#4
\arrowlength=#5
\ifnum \arrowtype<0
    \ifnum \run=0
        \advance \ypos by-\arrowlength
    \else
        \tempcounta \arrowlength
        \multiply \tempcounta by\rise
        \divide \tempcounta by\run
        \ifnum\run>0
            \advance \xpos by\arrowlength
            \advance \ypos by\tempcounta
        \else
            \advance \xpos by-\arrowlength
            \advance \ypos by-\tempcounta
        \fi
    \fi
    \multiply \arrowtype by-1
    \multiply \rise by-1
    \multiply \run by-1
\fi
\ifcase \arrowtype
\or \put(\xpos,\ypos){\vector(\run,\rise){\arrowlength}}%
\or \put(\xpos,\ypos){\mvector(\run,\rise)\arrowlength}%
\or \put(\xpos,\ypos){\evector(\run,\rise){\arrowlength}}%
\fi\fi\fi
}}
 
\def\putsplitvector(#1,#2)#3#4{%%
\xpos #1
\ypos #2
\arrowtype #4
\halflength #3
\arrowlength #3
\gap 140
\advance \halflength by-\gap
\divide \halflength by2
\ifnum\arrowtype>0
   \ifcase \arrowtype
   \or \put(\xpos,\ypos){\line(0,-1){\halflength}}%
       \advance\ypos by-\halflength
       \advance\ypos by-\gap
       \put(\xpos,\ypos){\vector(0,-1){\halflength}}%
   \or \put(\xpos,\ypos){\line(0,-1)\halflength}%
       \put(\xpos,\ypos){\vector(0,-1)3}%
       \advance\ypos by-\halflength
       \advance\ypos by-\gap
       \put(\xpos,\ypos){\vector(0,-1){\halflength}}%
   \or \put(\xpos,\ypos){\line(0,-1)\halflength}%
       \advance\ypos by-\halflength
       \advance\ypos by-\gap
       \put(\xpos,\ypos){\evector(0,-1){\halflength}}%
   \fi
\else \arrowtype=-\arrowtype
   \ifcase\arrowtype
   \or \advance \ypos by-\arrowlength
       \put(\xpos,\ypos){\line(0,1){\halflength}}%
       \advance\ypos by\halflength
       \advance\ypos by\gap
       \put(\xpos,\ypos){\vector(0,1){\halflength}}%
   \or \advance \ypos by-\arrowlength
       \put(\xpos,\ypos){\line(0,1)\halflength}%
       \put(\xpos,\ypos){\vector(0,1)3}%
       \advance\ypos by\halflength
       \advance\ypos by\gap
       \put(\xpos,\ypos){\vector(0,1){\halflength}}%
   \or \advance \ypos by-\arrowlength
       \put(\xpos,\ypos){\line(0,1)\halflength}%
       \advance\ypos by\halflength
       \advance\ypos by\gap
       \put(\xpos,\ypos){\evector(0,1){\halflength}}%
   \fi
\fi
}
 
\def\putmorphism(#1)(#2,#3)[#4`#5`#6]#7#8#9{{%
\run #2
\rise #3
\ifnum\rise=0
  \puthmorphism(#1)[#4`#5`#6]{#7}{#8}#9%
\else\ifnum\run=0
  \putvmorphism(#1)[#4`#5`#6]{#7}{#8}#9%
\else
\setpos(#1)%
\arrowlength #7
\arrowtype #8
\ifnum\run=0
\else\ifnum\rise=0
\else
\ifnum\run>0
    \coefa=1
\else
   \coefa=-1
\fi
\ifnum\arrowtype>0
   \coefb=0
   \coefc=-1
\else
   \coefb=\coefa
   \coefc=1
   \arrowtype=-\arrowtype
\fi
\width=2
\multiply \width by\run
\divide \width by\rise
\ifnum \width<0  \width=-\width\fi
\advance\width by60
\if l#9 \width=-\width\fi
\putbox(\xpos,\ypos){#4}%            %node 1
{\multiply \coefa by\arrowlength%      %node 2
\advance\xpos by\coefa
\multiply \coefa by\rise
\divide \coefa by\run
\advance \ypos by\coefa
\putbox(\xpos,\ypos){#5} }%
{\multiply \coefa by\arrowlength%      %label
\divide \coefa by2
\advance \xpos by\coefa
\advance \xpos by\width
\multiply \coefa by\rise
\divide \coefa by\run
\advance \ypos by\coefa
\if l#9%
   \putrbox(\xpos,\ypos){#6}%
\else\if r#9%
   \putlbox(\xpos,\ypos){#6}%
\fi\fi }%
{\multiply \rise by-\coefc%             %arrow
\multiply \run by-\coefc
\multiply \coefb by\arrowlength
\advance \xpos by\coefb
\multiply \coefb by\rise
\divide \coefb by\run
\advance \ypos by\coefb
\multiply \coefc by70
\advance \ypos by\coefc
\multiply \coefc by\run
\divide \coefc by\rise
\advance \xpos by\coefc
\multiply \coefa by140
\multiply \coefa by\run
\divide \coefa by\rise
\advance \arrowlength by\coefa
\ifcase\arrowtype
\or \put(\xpos,\ypos){\vector(\run,\rise){\arrowlength}}%
\or \put(\xpos,\ypos){\mvector(\run,\rise){\arrowlength}}%
\or \put(\xpos,\ypos){\evector(\run,\rise){\arrowlength}}%
\fi}\fi\fi\fi\fi}}

\newcount\numbdashes \newcount\lengthdash \newcount\increment
 
\def\howmanydashes{% Actually returns both number and length
\numbdashes=\arrowlength \lengthdash=40
\divide\numbdashes by \lengthdash
\lengthdash=\arrowlength
\divide\lengthdash by \numbdashes
\increment=\lengthdash
\multiply\lengthdash by 3
\divide\lengthdash by 5
}
 
\def\putdashvector(#1)(#2,#3)#4#5{%
\ifnum#3=0 \putdashhvector(#1){#4}#5
\else
\ifnum#2=0
\putdashvvector(#1){#4}#5\fi\fi}
 
\def\putdashhvector(#1,#2)#3#4{{%
\arrowlength=#3 \howmanydashes
\multiput(#1,#2)(\increment,0){\numbdashes}%
{\vrule height .4pt width \lengthdash\unitlength}
\arrowtype=#4 \xpos=#1
\ifnum\arrowtype<0 \advance\arrowtype by 7 \fi
\ifcase\arrowtype
\or \advance\xpos by 10
    \put(\xpos,#2){\vector(-1,0){\lengthdash}}
    \advance\xpos by 40
    \put(\xpos,#2){\vector(-1,0){\lengthdash}}
\or \advance \xpos by 10
    \put(\xpos,#2){\vector(-1,0){\lengthdash}}
    \advance\xpos by  \arrowlength
    \advance\xpos by  -50
    \put(\xpos,#2){\vector(-1,0){\lengthdash}}
\or \advance\xpos by 10
    \put(\xpos,#2){\vector(-1,0){\lengthdash}}
\or \advance\xpos by \arrowlength
    \advance\xpos by -\lengthdash
    \put(\xpos,#2){\vector(1,0){\lengthdash}}
\or {\advance\xpos by 10
    \put(\xpos,#2){\vector(1,0){\lengthdash}}}
    \advance\xpos by \arrowlength
    \advance\xpos by -\lengthdash
    \put(\xpos,#2){\vector(1,0){\lengthdash}}
\or \advance\xpos by \arrowlength
    \advance\xpos by -\lengthdash
    \put(\xpos,#2){\vector(1,0){\lengthdash}}
    \advance\xpos by -40
    \put(\xpos,#2){\vector(1,0){\lengthdash}}
   \fi
}}
 
\def\putdashvvector(#1,#2)#3#4{{%
\arrowlength=#3 \howmanydashes
\ypos=#2 \advance\ypos by -\arrowlength
\multiput(#1,#2)(0,\increment){\numbdashes}%
    {\vrule width .4pt height \lengthdash\unitlength}
\arrowtype=#4 \ypos=#2
\ifnum\arrowtype<0 \advance\arrowtype by 7 \fi
\ifcase\arrowtype
\or \advance\ypos by \arrowlength \advance\ypos by -40
    \put(#1,\ypos){\vector(0,1){\lengthdash}}
    \advance\ypos by -40
    \put(#1,\ypos){\vector(0,1){\lengthdash}}
\or \advance\ypos by 10
    \put(#1,\ypos){\vector(0,1){\lengthdash}}
    \advance\ypos by \arrowlength \advance\ypos by -40
    \put(#1,\ypos){\vector(0,1){\lengthdash}}
\or \advance\ypos by \arrowlength \advance\ypos by -40
    \put(#1,\ypos){\vector(0,1){\lengthdash}}
\or \advance\ypos by 10
    \put(#1,\ypos){\vector(0,-1){\lengthdash}}
\or \advance\ypos by 10
    \put(#1,\ypos){\vector(0,-1){\lengthdash}}
    \advance\ypos by \arrowlength \advance\ypos by -40
    \put(#1,\ypos){\vector(0,-1){\lengthdash}}
\or \advance\ypos by 10
    \put(#1,\ypos){\vector(0,-1){\lengthdash}}
    \advance\ypos by 40
    \put(#1,\ypos){\vector(0,-1){\lengthdash}}
\fi
}}
 
\def\puthmorphism(#1,#2)[#3`#4`#5]#6#7#8{{%
\xpos #1
\ypos #2
\width #6
\arrowlength #6
\arrowtype=#7
\putbox(\xpos,\ypos){#3\vphantom{#4}}%
{\advance \xpos by\arrowlength
\putbox(\xpos,\ypos){\vphantom{#3}#4}}%
\horsize{\tempcounta}{#3}%
\horsize{\tempcountb}{#4}%
\divide \tempcounta by2
\divide \tempcountb by2
\advance \tempcounta by30
\advance \tempcountb by30
\advance \xpos by\tempcounta
\advance \arrowlength by-\tempcounta
\advance \arrowlength by-\tempcountb
\putvector(\xpos,\ypos)(1,0)\arrowlength\arrowtype
\divide \arrowlength by2
\advance \xpos by\arrowlength
\vertsize{\tempcounta}{#5}%
\divide\tempcounta by2
\advance \tempcounta by20
\if a#8 %
   \advance \ypos by\tempcounta
   \putbox(\xpos,\ypos){#5}%
\else
   \advance \ypos by-\tempcounta
   \putbox(\xpos,\ypos){#5}%
\fi}}
 
\def\putvmorphism(#1,#2)[#3`#4`#5]#6#7#8{{%
\xpos #1
\ypos #2
\arrowlength #6
\arrowtype #7
\settowidth{\xlen}{$#5$}%
\putbox(\xpos,\ypos){#3}%
{\advance \ypos by-\arrowlength
\putbox(\xpos,\ypos){#4}}%
{\advance\arrowlength by-140
\advance \ypos by-70
\ifdim\xlen>0pt
   \if m#8%
      \putsplitvector(\xpos,\ypos)\arrowlength\arrowtype
   \else
   \putvector(\xpos,\ypos)(0,-1)\arrowlength\arrowtype
   \fi
\else
   \putvector(\xpos,\ypos)(0,-1)\arrowlength\arrowtype
\fi}%
\ifdim\xlen>0pt
   \divide \arrowlength by2
   \advance\ypos by-\arrowlength
   \if l#8%
      \advance \xpos by-40
      \putrbox(\xpos,\ypos){#5}%
   \else\if r#8%
      \advance \xpos by40
      \putlbox(\xpos,\ypos){#5}%
   \else
      \putbox(\xpos,\ypos){#5}%
   \fi\fi
\fi
}}
 
\def\putsquarep<#1>(#2)[#3;#4`#5`#6`#7]{{%
\setsqparms[#1]%
\setpos(#2)%
\settokens`#3`%
\puthmorphism(\xpos,\ypos)[\tokenc`\tokend`{#7}]{\width}{\arrowtyped}b%
\advance\ypos by \height
\puthmorphism(\xpos,\ypos)[\tokena`\tokenb`{#4}]{\width}{\arrowtypea}a%
\putvmorphism(\xpos,\ypos)[``{#5}]{\height}{\arrowtypeb}l%
\advance\xpos by \width
\putvmorphism(\xpos,\ypos)[``{#6}]{\height}{\arrowtypec}r%
}}
 
\def\putsquare{\@ifnextchar <{\putsquarep}{\putsquarep%
   <\arrowtypea`\arrowtypeb`\arrowtypec`\arrowtyped;\width`\height>}}
\def\square{\@ifnextchar< {\squarep}{\squarep
   <\arrowtypea`\arrowtypeb`\arrowtypec`\arrowtyped;\width`\height>}}
\def\squarep<#1>[#2`#3`#4`#5;#6`#7`#8`#9]{{%       %     #2------>#3
\setsqparms[#1]%                                   %      |       |
\diagram%                                          %      |       |
\putsquarep<\arrowtypea`\arrowtypeb`\arrowtypec`%  %    #7|       |#8
\arrowtyped;\width`\height>%                       %      |       |
(0,0)[#2`#3`#4`{#5};#6`#7`#8`{#9}]%                %      |       |
\enddiagram%                                       %      v       v
}}                                                 %     #4------>#5
\def\putptrianglep<#1>(#2,#3)[#4`#5`#6;#7`#8`#9]{{%
\settriparms[#1]%
\xpos=#2 \ypos=#3
\advance\ypos by \height
\puthmorphism(\xpos,\ypos)[#4`#5`{#7}]{\height}{\arrowtypea}a%
\putvmorphism(\xpos,\ypos)[`#6`{#8}]{\height}{\arrowtypeb}l%
\advance\xpos by\height
\putmorphism(\xpos,\ypos)(-1,-1)[``{#9}]{\height}{\arrowtypec}r%
}}
 
\def\putptriangle{\@ifnextchar <{\putptrianglep}{\putptrianglep
   <\arrowtypea`\arrowtypeb`\arrowtypec;\height>}}
\def\ptriangle{\@ifnextchar <{\ptrianglep}{\ptrianglep
   <\arrowtypea`\arrowtypeb`\arrowtypec;\height>}}
\def\ptrianglep<#1>[#2`#3`#4;#5`#6`#7]{{%%    %      #2----->#3
\settriparms[#1]%                             %      |      /
\diagram%                                     %      |     /
\putptrianglep<\arrowtypea`\arrowtypeb`%      %    #6|    /#7
\arrowtypec;\height>%                         %      |   /
(0,0)[#2`#3`#4;#5`#6`{#7}]%                   %      |  /
\enddiagram%%                                 %      v v
}}                                            %      #4
 
\def\putqtrianglep<#1>(#2,#3)[#4`#5`#6;#7`#8`#9]{{%
\settriparms[#1]%
\xpos=#2 \ypos=#3
\advance\ypos by\height
\puthmorphism(\xpos,\ypos)[#4`#5`{#7}]{\height}{\arrowtypea}a%
\putmorphism(\xpos,\ypos)(1,-1)[``{#8}]{\height}{\arrowtypeb}l%
\advance\xpos by\height
\putvmorphism(\xpos,\ypos)[`#6`{#9}]{\height}{\arrowtypec}r%
}}
 
\def\putqtriangle{\@ifnextchar <{\putqtrianglep}{\putqtrianglep
   <\arrowtypea`\arrowtypeb`\arrowtypec;\height>}}
\def\qtriangle{\@ifnextchar <{\qtrianglep}{\qtrianglep
   <\arrowtypea`\arrowtypeb`\arrowtypec;\height>}}
\def\qtrianglep<#1>[#2`#3`#4;#5`#6`#7]{{%%    %        #2----->#3
\settriparms[#1]%                             %         \      |
\width=\height                                %          \     |
\diagram%                                     %         #6\    |#7
\putqtrianglep<\arrowtypea`\arrowtypeb`%      %            \   |
\arrowtypec;\height>%                         %             \  |
(0,0)[#2`#3`#4;#5`#6`{#7}]%                   %              v v
\enddiagram%%                                 %               #4
}}
 
\def\putdtrianglep<#1>(#2,#3)[#4`#5`#6;#7`#8`#9]{{%
\settriparms[#1]%
\xpos=#2 \ypos=#3
\puthmorphism(\xpos,\ypos)[#5`#6`{#9}]{\height}{\arrowtypec}b%
\advance\xpos by \height \advance\ypos by\height
\putmorphism(\xpos,\ypos)(-1,-1)[``{#7}]{\height}{\arrowtypea}l%
\putvmorphism(\xpos,\ypos)[#4``{#8}]{\height}{\arrowtypeb}r%
}}
 
\def\putdtriangle{\@ifnextchar <{\putdtrianglep}{\putdtrianglep
   <\arrowtypea`\arrowtypeb`\arrowtypec;\height>}}
\def\dtriangle{\@ifnextchar <{\dtrianglep}{\dtrianglep
   <\arrowtypea`\arrowtypeb`\arrowtypec;\height>}}
\def\dtrianglep<#1>[#2`#3`#4;#5`#6`#7]{{%%    %                  / |
\settriparms[#1]%                             %                 /  |
\width=\height                                %              #5/   |#6
\diagram%                                     %               /    |
\putdtrianglep<\arrowtypea`\arrowtypeb`%      %              /     |
\arrowtypec;\height>%                         %             v      v
(0,0)[#2`#3`#4;#5`#6`{#7}]%                   %            #3----->#4
\enddiagram%%                                 %                #7
}}
 
\def\putbtrianglep<#1>(#2,#3)[#4`#5`#6;#7`#8`#9]{{%
\settriparms[#1]%
\xpos=#2 \ypos=#3
\puthmorphism(\xpos,\ypos)[#5`#6`{#9}]{\height}{\arrowtypec}b%
\advance\ypos by\height
\putmorphism(\xpos,\ypos)(1,-1)[``{#8}]{\height}{\arrowtypeb}r%
\putvmorphism(\xpos,\ypos)[#4``{#7}]{\height}{\arrowtypea}l%
}}
 
\def\putbtriangle{\@ifnextchar <{\putbtrianglep}{\putbtrianglep
   <\arrowtypea`\arrowtypeb`\arrowtypec;\height>}}
\def\btriangle{\@ifnextchar <{\btrianglep}{\btrianglep
   <\arrowtypea`\arrowtypeb`\arrowtypec;\height>}}
\def\btrianglep<#1>[#2`#3`#4;#5`#6`#7]{{%%   %              | \
\settriparms[#1]%                            %              |  \
\width=\height                               %            #5|   \#6
\diagram%                                    %              |    \
\putbtrianglep<\arrowtypea`\arrowtypeb`%     %              |     \
\arrowtypec;\height>%                        %              v      v
(0,0)[#2`#3`#4;#5`#6`{#7}]%                  %              #3----->#4
\enddiagram%%                                %                 #7
}}
 
\def\putAtrianglep<#1>(#2,#3)[#4`#5`#6;#7`#8`#9]{{%
\settriparms[#1]%
\xpos=#2 \ypos=#3
{\multiply \height by2
\puthmorphism(\xpos,\ypos)[#5`#6`{#9}]{\height}{\arrowtypec}b}%
\advance\xpos by\height \advance\ypos by\height
\putmorphism(\xpos,\ypos)(-1,-1)[#4``{#7}]{\height}{\arrowtypea}l%
\putmorphism(\xpos,\ypos)(1,-1)[``{#8}]{\height}{\arrowtypeb}r%
}}
 
\def\putAtriangle{\@ifnextchar <{\putAtrianglep}{\putAtrianglep
   <\arrowtypea`\arrowtypeb`\arrowtypec;\height>}}
\def\Atriangle{\@ifnextchar <{\Atrianglep}{\Atrianglep
   <\arrowtypea`\arrowtypeb`\arrowtypec;\height>}}
\def\Atrianglep<#1>[#2`#3`#4;#5`#6`#7]{{%%         %         /   \
\settriparms[#1]%                                  %        /     \
\width=\height                                     %     #5/       \#6
\diagram%                                          %      /         \
\putAtrianglep<\arrowtypea`\arrowtypeb`%           %     /           \
\arrowtypec;\height>%                              %    v             v
(0,0)[#2`#3`#4;#5`#6`{#7}]%                        %   #3------------>#4
\enddiagram%%                                      %          #7
}}
 
\def\putAtrianglepairp<#1>(#2)[#3;#4`#5`#6`#7`#8]{{%
\settripairparms[#1]%
\setpos(#2)%
\settokens`#3`%
\puthmorphism(\xpos,\ypos)[\tokenb`\tokenc`{#7}]{\height}{\arrowtyped}b%
\advance\xpos by\height
\puthmorphism(\xpos,\ypos)[\phantom{\tokenc}`\tokend`{#8}]%
{\height}{\arrowtypee}b%
\advance\ypos by\height
\putmorphism(\xpos,\ypos)(-1,-1)[\tokena``{#4}]{\height}{\arrowtypea}l%
\putvmorphism(\xpos,\ypos)[``{#5}]{\height}{\arrowtypeb}m%
\putmorphism(\xpos,\ypos)(1,-1)[``{#6}]{\height}{\arrowtypec}r%
}}
 
\def\putAtrianglepair{\@ifnextchar <{\putAtrianglepairp}{\putAtrianglepairp%
   <\arrowtypea`\arrowtypeb`\arrowtypec`\arrowtyped`\arrowtypee;\height>}}
\def\Atrianglepair{\@ifnextchar <{\Atrianglepairp}{\Atrianglepairp%
   <\arrowtypea`\arrowtypeb`\arrowtypec`\arrowtyped`\arrowtypee;\height>}}
 
\def\Atrianglepairp<#1>[#2;#3`#4`#5`#6`#7]{{%           %  #2a
\settripairparms[#1]%                         %           / | \
\settokens`#2`%                               %          /  |  \
\width=\height                                %       #3/  #4   \#5
\diagram%                                     %        /    |    \
\putAtrianglepairp                            %       /     |     \
<\arrowtypea`\arrowtypeb`\arrowtypec`%        %      v      v      v
\arrowtyped`\arrowtypee;\height>%             %     #2b---->#2c---->#2d
(0,0)[{#2};#3`#4`#5`#6`{#7}]%                 %         #6     #7
\enddiagram%%
}}
 
\def\putVtrianglep<#1>(#2,#3)[#4`#5`#6;#7`#8`#9]{{%
\settriparms[#1]%
\xpos=#2 \ypos=#3
\advance\ypos by\height
{\multiply\height by2
\puthmorphism(\xpos,\ypos)[#4`#5`{#7}]{\height}{\arrowtypea}a}%
\putmorphism(\xpos,\ypos)(1,-1)[`#6`{#8}]{\height}{\arrowtypeb}l%
\advance\xpos by\height
\advance\xpos by\height
\putmorphism(\xpos,\ypos)(-1,-1)[``{#9}]{\height}{\arrowtypec}r%
}}
 
\def\putVtriangle{\@ifnextchar <{\putVtrianglep}{\putVtrianglep
   <\arrowtypea`\arrowtypeb`\arrowtypec;\height>}}
\def\Vtriangle{\@ifnextchar <{\Vtrianglep}{\Vtrianglep
   <\arrowtypea`\arrowtypeb`\arrowtypec;\height>}}
\def\Vtrianglep<#1>[#2`#3`#4;#5`#6`#7]{{%%     %        #2------------->#3
\settriparms[#1]%                              %         \             /
\width=\height                                 %          \           /
\diagram%                                      %         #6\         /#7
\putVtrianglep<\arrowtypea`\arrowtypeb`%       %            \       /
\arrowtypec;\height>%                          %             \     /
(0,0)[#2`#3`#4;#5`#6`{#7}]%                    %              v   v
\enddiagram%%                                  %               #4
}}
 
\def\putVtrianglepairp<#1>(#2)[#3;#4`#5`#6`#7`#8]{{
\settripairparms[#1]%
\setpos(#2)%
\settokens`#3`%
\advance\ypos by\height
\putmorphism(\xpos,\ypos)(1,-1)[`\tokend`{#6}]{\height}{\arrowtypec}l%
\puthmorphism(\xpos,\ypos)[\tokena`\tokenb`{#4}]{\height}{\arrowtypea}a%
\advance\xpos by\height
\puthmorphism(\xpos,\ypos)[\phantom{\tokenb}`\tokenc`{#5}]%
{\height}{\arrowtypeb}a%
\putvmorphism(\xpos,\ypos)[``{#7}]{\height}{\arrowtyped}m%
\advance\xpos by\height
\putmorphism(\xpos,\ypos)(-1,-1)[``{#8}]{\height}{\arrowtypee}r%
}}
 
\def\putVtrianglepair{\@ifnextchar <{\putVtrianglepairp}{\putVtrianglepairp%
    <\arrowtypea`\arrowtypeb`\arrowtypec`\arrowtyped`\arrowtypee;\height>}}
\def\Vtrianglepair{\@ifnextchar <{\Vtrianglepairp}{\Vtrianglepairp%
    <\arrowtypea`\arrowtypeb`\arrowtypec`\arrowtyped`\arrowtypee;\height>}}
\def\Vtrianglepairp<#1>[#2;#3`#4`#5`#6`#7]{{%  %  #2a---->#2b---->#2c
\settripairparms[#1]%                          %   \      |      /
\settokens`#2`%                                %    \     |     /
\diagram%                                      %   #5\   #6    /#7
\putVtrianglepairp                             %      \   |   /
<\arrowtypea`\arrowtypeb`\arrowtypec`%         %       \  |  /
\arrowtyped`\arrowtypee;\height>%              %        v v v
(0,0)[{#2};#3`#4`#5`#6`{#7}]%                  %         #2d
\enddiagram%%
}}

\def\putCtrianglep<#1>(#2,#3)[#4`#5`#6;#7`#8`#9]{{%
\settriparms[#1]%
\xpos=#2 \ypos=#3
\advance\ypos by\height
\putmorphism(\xpos,\ypos)(1,-1)[``{#9}]{\height}{\arrowtypec}l%
\advance\xpos by\height
\advance\ypos by\height
\putmorphism(\xpos,\ypos)(-1,-1)[#4`#5`{#7}]{\height}{\arrowtypea}l%
{\multiply\height by 2
\putvmorphism(\xpos,\ypos)[`#6`{#8}]{\height}{\arrowtypeb}r}%
}}
 
\def\putCtriangle{\@ifnextchar <{\putCtrianglep}{\putCtrianglep
    <\arrowtypea`\arrowtypeb`\arrowtypec;\height>}}
\def\Ctriangle{\@ifnextchar <{\Ctrianglep}{\Ctrianglep
    <\arrowtypea`\arrowtypeb`\arrowtypec;\height>}}
\def\Ctrianglep<#1>[#2`#3`#4;#5`#6`#7]{{%%   %                / |
\settriparms[#1]%                            %             #5/  |
\width=\height                               %              /   |
\diagram%                                    %             v    |
\putCtrianglep<\arrowtypea`\arrowtypeb`%     %           #3     |#6
\arrowtypec;\height>%                        %             \    |
(0,0)[#2`#3`#4;#5`#6`{#7}]%                  %            #7\   |
\enddiagram%%                                %               \  |
}}                                           %                v v
\def\putDtrianglep<#1>(#2,#3)[#4`#5`#6;#7`#8`#9]{{%
\settriparms[#1]%
\xpos=#2 \ypos=#3
\advance\xpos by\height \advance\ypos by\height
\putmorphism(\xpos,\ypos)(-1,-1)[``{#9}]{\height}{\arrowtypec}r%
\advance\xpos by-\height \advance\ypos by\height
\putmorphism(\xpos,\ypos)(1,-1)[`#5`{#8}]{\height}{\arrowtypeb}r%
{\multiply\height by 2
\putvmorphism(\xpos,\ypos)[#4`#6`{#7}]{\height}{\arrowtypea}l}%
}}
 
\def\putDtriangle{\@ifnextchar <{\putDtrianglep}{\putDtrianglep
    <\arrowtypea`\arrowtypeb`\arrowtypec;\height>}}
\def\Dtriangle{\@ifnextchar <{\Dtrianglep}{\Dtrianglep
   <\arrowtypea`\arrowtypeb`\arrowtypec;\height>}}
\def\Dtrianglep<#1>[#2`#3`#4;#5`#6`#7]{{%%  %          | \
\settriparms[#1]%                           %          |  \#6
\width=\height                              %          |   \
\diagram%                                   %          |    v
\putDtrianglep<\arrowtypea`\arrowtypeb`%    %        #5|    #3
\arrowtypec;\height>%                       %          |    /
(0,0)[#2`#3`#4;#5`#6`{#7}]%                 %          |   /#7
\enddiagram%%                               %          |  /
}}                                          %          v v
\def\setrecparms[#1`#2]{\width=#1 \height=#2}%
\def\recursep<#1`#2>[#3;#4`#5`#6`#7`#8]{{\m@th
\width=#1 \height=#2
\settokens`#3`
\settowidth{\tempdimen}{$\tokena$}
\ifdim\tempdimen=0pt
  \savebox{\tempboxa}{\hbox{$\tokenb$}}%
  \savebox{\tempboxb}{\hbox{$\tokend$}}%
  \savebox{\tempboxc}{\hbox{$#6$}}%
\else
  \savebox{\tempboxa}{\hbox{$\hbox{$\tokena$}\times\hbox{$\tokenb$}$}}%
  \savebox{\tempboxb}{\hbox{$\hbox{$\tokena$}\times\hbox{$\tokend$}$}}%
  \savebox{\tempboxc}{\hbox{$\hbox{$\tokena$}\times\hbox{$#6$}$}}%
\fi
\ypos=\height
\divide\ypos by 2
\xpos=\ypos
\advance\xpos by \width
\bfig
\putCtrianglep<-1`1`1;\ypos>(0,0)[`\tokenc`;#5`#6`{#7}]%
\puthmorphism(\ypos,0)[\tokend`\usebox{\tempboxb}`{#8}]{\width}{-1}b%
\puthmorphism(\ypos,\height)[\tokenb`\usebox{\tempboxa}`{#4}]{\width}{-1}a%
\advance\ypos by \width
\putvmorphism(\ypos,\height)[``\usebox{\tempboxc}]{\height}1r%
\efig
}}
 
\def\recurse{\@ifnextchar <{\recursep}{\recursep<\width`\height>}}
 
\def\puttwohmorphisms(#1,#2)[#3`#4;#5`#6]#7#8#9{{%
\puthmorphism(#1,#2)[#3`#4`]{#7}0a
\ypos=#2
\advance\ypos by 20
\puthmorphism(#1,\ypos)[\phantom{#3}`\phantom{#4}`#5]{#7}{#8}a
\advance\ypos by -40
\puthmorphism(#1,\ypos)[\phantom{#3}`\phantom{#4}`#6]{#7}{#9}b
}}
 
\def\puttwovmorphisms(#1,#2)[#3`#4;#5`#6]#7#8#9{{%
\putvmorphism(#1,#2)[#3`#4`]{#7}0a
\xpos=#1
\advance\xpos by -20
\putvmorphism(\xpos,#2)[\phantom{#3}`\phantom{#4}`#5]{#7}{#8}l
\advance\xpos by 40
\putvmorphism(\xpos,#2)[\phantom{#3}`\phantom{#4}`#6]{#7}{#9}r
}}
 
\def\puthcoequalizer(#1)[#2`#3`#4;#5`#6`#7]#8#9{{%
\setpos(#1)%
\puttwohmorphisms(\xpos,\ypos)[#2`#3;#5`#6]{#8}11%
\advance\xpos by #8
\puthmorphism(\xpos,\ypos)[\phantom{#3}`#4`#7]{#8}1{#9}
}}
 
\def\putvcoequalizer(#1)[#2`#3`#4;#5`#6`#7]#8#9{{%
\setpos(#1)%
\puttwovmorphisms(\xpos,\ypos)[#2`#3;#5`#6]{#8}11%
\advance\ypos by -#8
\putvmorphism(\xpos,\ypos)[\phantom{#3}`#4`#7]{#8}1{#9}
}}
 
\def\putthreehmorphisms(#1)[#2`#3;#4`#5`#6]#7(#8)#9{{%
\setpos(#1) \settypes(#8)
\if a#9 %
     \vertsize{\tempcounta}{#5}%
     \vertsize{\tempcountb}{#6}%
     \ifnum \tempcounta<\tempcountb \tempcounta=\tempcountb \fi
\else
     \vertsize{\tempcounta}{#4}%
     \vertsize{\tempcountb}{#5}%
     \ifnum \tempcounta<\tempcountb \tempcounta=\tempcountb \fi
\fi
\advance \tempcounta by 60
\puthmorphism(\xpos,\ypos)[#2`#3`#5]{#7}{\arrowtypeb}{#9}
\advance\ypos by \tempcounta
\puthmorphism(\xpos,\ypos)[\phantom{#2}`\phantom{#3}`#4]{#7}{\arrowtypea}{#9}
\advance\ypos by -\tempcounta \advance\ypos by -\tempcounta
\puthmorphism(\xpos,\ypos)[\phantom{#2}`\phantom{#3}`#6]{#7}{\arrowtypec}{#9}
}}
 
\def\setarrowtoks[#1`#2`#3`#4`#5`#6]{%
\def\toka{#1}
\def\tokb{#2}
\def\tokc{#3}
\def\tokd{#4}
\def\toke{#5}
\def\tokf{#6}
}
\def\hex{\@ifnextchar <{\hexp}{\hexp<1000`400>}}
\def\hexp<#1`#2>[#3`#4`#5`#6`#7`#8;#9]{%
\setarrowtoks[#9]
\yext=#2 \advance \yext by #2
\xext=#1 \advance\xext by \yext
\bfig
\putCtriangle<-1`0`1;#2>(0,0)[`#5`;\tokb``\tokd]
\xext=#1 \yext=#2 \advance \yext by #2
\putsquare<1`0`0`1;\xext`\yext>(#2,0)[#3`#4`#7`#8;\toka```\tokf]
\advance \xext by #2
\putDtriangle<0`1`-1;#2>(\xext,0)[`#6`;`\tokc`\toke]
\efig
}
\makeatother